\UseRawInputEncoding
\documentclass[utf8]{FrontiersinHarvard} 

\usepackage{times}
\usepackage{amsmath}
\usepackage{xcolor,multicol,cuted}
\usepackage{ulem}
\usepackage{url,hyperref,lineno,microtype,subcaption}
\usepackage[onehalfspacing]{setspace}

\def\ltsima{$\; \buildrel < \over \sim \;$}
\def\simlt{\lower.5ex\hbox{\ltsima}}
\def\gtsima{$\; \buildrel > \over \sim \;$}
\def\simgt{\lower.5ex\hbox{\gtsima}}

\newcommand{\msun}{{\rm\,\thinspace M_\odot}}

\newcommand{\mm}{~{\rm m m} }

\newcommand{\kms}{~{\rm km~ s^{-1}} } 
\newcommand{\mic}{$\mu$m}

\newcommand{\arcsec}{\mbox{$^{\prime \prime}$}}

\def\keyFont{\fontsize{8}{11}\helveticabold }
\def\firstAuthorLast{Maury {et~al.}}
\def\Authors{Anaëlle Maury\,$^{1,2,*}$, Patrick Hennebelle\,$^{1}$ and Josep Miquel Girart\,$^{3,4}$}

\def\Address{$^{1}$CEA/DRF/IRFU Astrophysics department, Paris-Saclay University, UMR AIM, Gif-sur-Yvette, France \\
$^{2}$Harvard-Smithsonian Center for Astrophysics, Cambridge, USA \\
$^{3}$
Institut de Ci\`encies de l'Espai (ICE), CSIC,
Can Magrans s/n, E-08193 Cerdanyola del Vall\`es,  Catalonia, Spain
\\$^{4}$ 
Institut d'Estudis Espacials de Catalunya (IEEC), E-08034 Barcelona, Catalonia, Spain}

\def\corrAuthor{CEA/DRF/IRFU/DAp, Orme des Merisiers batiment 709, 91191 Gif-sur-Yvette, France}
\def\corrEmail{anaelle.maury@cea.fr}

\begin{document}
\firstpage{1}
\title[Magnetic fields in embedded protostars]{Recent progress with observations and models to characterize the magnetic fields from star-forming cores to protostellar disks} 

\author[\firstAuthorLast ]{\Authors} 
\address{} 
\correspondance{} 

\extraAuth{}

\maketitle


\begin{abstract}
In this review paper, we aim at providing a global outlook on the progresses made in the recent years to characterize the role of magnetic fields during the embedded phases of the star formation process. Thanks to the development of observational capabilities and the parallel progress in numerical models capturing most of the important physics at work during star formation, it has recently become possible to confront detailed predictions of magnetized models to observational properties of the youngest protostars. We provide an overview of the most important consequences when adding magnetic fields to state-of-the-art models of protostellar formation, emphasizing their role  to shape the resulting star(s) and their disk(s). We discuss the importance of magnetic field coupling to set the efficiency of magnetic processes, and provide a review of observational works putting constraints on the two main agents responsible for the coupling in star-forming cores: dust grains and ionized gas. We recall the physical processes and observational methods allowing to trace the magnetic field topology and its intensity in embedded protostars, and review the main steps, success and limitations in comparing real observations to synthetic observations from the non-ideal MHD models. Finally, we discuss the main threads of observational evidence that suggest a key role of magnetic fields for star and disk formation, and propose a scenario solving the angular momentum for star formation, also highlighting the remaining tensions that exist between models and observations.
\tiny
 \keyFont{ \section{Keywords:} Star formation, magnetic fields, protostars, MHD modeling, polarization} 
\end{abstract}  

\section{Introduction}

The formation of stars takes place in filamentary molecular clouds, when the high-density interstellar medium partly collapses and fragments into bound starless dense cores. Depending on their properties, such cores can undergo further collapse to form a protostar that will accrete its circumstellar material (evolving through the Class~0, I, II and later stages) until reaching the main sequence.
Class~0 protostars are the first (proto)stellar objects, observed only $t \simlt 5 \times 10^4$~yr \citep{Andre93, Andre00, Evans09, Maury11} after their formation, while most of their mass is still in the form of a dense core/envelope collapsing onto the central protostellar embryo.
Embedded Class 0 and Class I protostars are also accreting objects, as most of the final stellar mass is assembled during those short phases. 
During this accretion phase, the circumstellar gas must redistribute most of its initial angular momentum outward or else centrifugal forces will soon balance gravity and prevent inflow, accretion, and the growth of the star.
This long-standing ``angular momentum problem'' was estimated to be quite severe \citep{Bodenheimer95}. Comparing observations of the specific angular momentum $J/M$ in star-forming clouds prior to contraction ($J/M \sim 10^{21} \,\rm{cm}^{2} \,\rm{s}^{-1}$, \citealt{Goodman93}) to the angular momentum contained in a typical solar-type star ($J/M \sim 10^{15} \,\rm{cm}^{2} \,\rm{s}^{-1}$) indicates that, during the brief accretion phase, the gas must transfer all but 1 millionth of its initial angular momentum out of the accretion flow \citep{Belloche13a}. 
Exactly how the circumstellar mass contained in a star-forming core is transferred to the forming star, but not its associated angular momentum, has been an active field of research in modern astronomy.

Wherever we have the means of observing them, magnetic fields are detected on nearly all scales and across the full spectrum of astrophysical environments: from our own Earth \citep{Russell91} to stars \citep{Donati09}, the Milky Way \citep{Wielebinski05} and cosmological structures \citep{Kunze13}. 
As regards star-formation processes, magnetic fields provide a mechanism for launching and collimating outflows winds and jets \citep{Blandford82}, which are routinely observed around young stellar objects (YSOs), and magnetic fields of typical strengths 10--100 $\mu$G are threading nearly all star-forming clouds \citep{Crutcher12}. Therefore, it is now widely accepted that most star-forming cores are magnetized to some level. However, it is only recently that the role of these magnetic fields could be investigated in details.

Indeed, in the past decade, numerical models of star formation have been gradually including most of the physical ingredients for a detailed description of protostellar evolution in the presence of magnetic fields, such as resistive MHD, radiative transfer and chemical networks, at all the relevant scales. 
Moreover, observational capabilities resolving the internal environment of star-forming cores, e.g., interferometry at (sub)millimeter wavelengths, have started to produce detailed maps of the gas and dust properties at the very small scales of protostellar cores, where material is accreted into a stellar embryo, stored in a disk, and ejected under the form of outflows and jets. The massive development of polarization capabilities on telescopes probing the cold Universe, such as on the SubMillimeter Array (SMA), then on the Atacama Large Millimeter Array (ALMA), and now also developed at the IRAM NOrthern Extended Millimeter Array (NOEMA), have produced sensitive observations of magnetic fields in protostellar cores with a great level of detail. These are used as tools to put unprecedented constraints on magnetic features the star-formation models should be able to reproduce \citep[e.g.][]{HullZhang19}.

Hence, the simultaneous major improvements of instrumental and computational facilities have opened an era of detailed confrontation between observed protostellar properties and magnetized models predictions.  Testing the detailed interplay of physics at work has allowed a major leap forward in our understanding of star and disk formation processes, and provide a new detailed scenario to describe the early stages of the formation of stars and their planetary systems. 
In this review, we present a synthetic description of the progresses made in the past decade regarding the properties and roles of magnetic fields in shaping young protostars, their envelopes, disks, and resulting stellar systems. 

\section{How magnetic field influences gravitational collapse: a theoretical overview}

This section is devoted to a review of the fundamental MHD processes relevant in the context of 
dense core collapse and disk formation. 

\subsection{The fluid equations}

The fluid equations are as follows.
$\rho$, $P$, $\mathbf{v}$, $\mathbf{B}$, and ${\rm \Phi}_g$ are, respectively, gas density,  pressure, velocity, magnetic field, and gravitational potential, while $\eta_{\rm O}$, $\eta_{\rm H}$, and $\eta_{\rm AD}$ are the Ohmic, Hall and ambipolar diffusivities.

\begin{equation}
{\partial \rho \over \partial t} + \nabla \cdot (\rho \mathbf{v}) = 0~,
\end{equation}
is the continuity equation which describes matter conservation.
\begin{eqnarray}
\label{navier}
\rho \left( {\partial \mathbf{v} \over \partial t} + (\mathbf{v}\cdot\nabla)\mathbf{v} \right)
\\ = -\nabla P  + {1 \over 4 \pi} (\nabla \times \mathbf{B}) \times \mathbf{B} - \rho \nabla {\rm \Phi}_g~ \nonumber
\\ = -\nabla \left( P + {B^2 \over 8 \pi} \right)+  {1 \over 4 \pi} \mathbf{B} . \nabla \mathbf{B} - \rho \nabla {\rm \Phi}_g~,
\nonumber
\end{eqnarray}
represents the momentum conservation. 
We see in particular that the Lorentz force can 
be written as a magnetic pressure and a magnetic tension. 
\begin{eqnarray}
\label{Eq:inductB}
{\partial \mathbf{B} \over \partial t} = \nabla \times (\mathbf{v \times B}) - {c^2 \over 4 \pi } \nabla 
\\ \times  \left\{\eta_{\rm O}\nabla \times \mathbf{B} + \eta_{\rm H}(\nabla \times \mathbf{B}) \times {\mathbf{B} \over B} \right. \nonumber 
\\ 
\left. + \eta_{\rm AD}{\mathbf{B} \over B} \times \left[(\nabla \times \mathbf{B}) \times {\mathbf{B} \over B} \right] \right\} ~,\nonumber
\end{eqnarray}
is the Maxwell-Faraday equation. The three non-ideal
MHD terms, namely the Ohm, Hall and ambipolar 
diffusion contributions are accounted for.

While this set of equations is complete (after the inclusion of an energy conservation equation, and an equation of state), it is nevertheless enlightening to also discuss the 
equation describing the  conservation of angular momentum. It is obtained by combining the 
azimuthal component of the momentum equation
with the continuity equation 
\citep[see for instance][]{2012A&A...543A.128J}.
The equation of conservation of angular momentum is
\begin{eqnarray}
{\partial_{t}\left(\rho rv_{\phi}\right) + \nabla \cdot r}
 \bigg[ \rho v_{\phi}{\bf v} 
 \\
\left.
 + \left(P + \frac{B^2}{8\pi}-\frac{g^2}{8 \pi G}\right){\bf e_{\phi}} 
   - \frac{B_{\phi}}{4\pi}{\bf B} + \frac{g_{\phi}}{4\pi G}{\bf g}\right] \nonumber = 0, \nonumber
\label{euler}
\end{eqnarray}
This equation reveals the existence of two 
torques able to transport the angular momentum,
namely $-r {B_{\phi}}{\bf B} / 4 \pi$, the 
magnetic torque and 
$r {g_{\phi}}{\bf g} / (4 \pi G)$, the gravitational 
torque. These two torques are playing a fundamental 
role during the collapse regarding the issue of 
angular momentum evolution. 
They have a similar expression, and in particular both require a toroidal 
and a poloidal fields. However, there is an important difference between the two 
because a toroidal magnetic field, $B_\phi$, can be produced in 
an axi-symmetric cloud, once the cloud is in rotation. On the contrary, 
a toroidal gravitational field, requires a non-axisymmetric density 
distribution such as a spiral wave.

\subsection{Magnetic support and magnetic compression: the pseudo-disk}
The simplest effect magnetic field has on a cloud is through magnetic pressure, which provides a support against gravity. Assuming field freezing within a cloud of radius $R$, threaded by magnetic field, $B$, we have that $\phi = \pi R^2 B$, the magnetic flux, is conserved during collapse. It is easy to calculate the ratio of magnetic over gravitational energy

\begin{eqnarray}
\label{emag}
{E_{\rm mag} \over E_{\rm grav} } \propto {R^3 B^2 \over G M^2 / R } \propto {1 \over G} \left( { \phi \over M}
\right)^2. 
\end{eqnarray}
Equation~\ref{emag} shows that the energy ratio stays constant, as long as spherical symmetry 
is maintained and that it is proportional to $\phi / M$, i.e. the ratio of the magnetic flux over the
cloud mass.  Obviously there is a critical value for $(M / \phi )_{cri†}$ above which 
magnetic field is strong enough to prevent gravitational collapse. 
The latter is expected to be such as $(M / \phi )_{cri†} \propto 1/\sqrt{G}$. 
A calibration has been performed by \citet{Mouschovias76} using exact equilibrium and, 
it has been inferred that 
\begin{eqnarray}
\left( {M \over \phi} \right) _{crit} = {1 \over \sqrt{G} } {c_1 \sqrt{5} \over 3 \pi },
\end{eqnarray}
where $c_1 \simeq 0.51$.
It is then common to define $\mu$, the mass-to-flux over critical mass-to-flux ratio. 
A value of $\mu=1$ indicates that magnetic field and gravity compensate while a value of
$\mu$ larger than 1, implies that magnetic field cannot prevent collapse to occur. Another commonly 
used closely related parameter has traditionally been called  $\lambda$ and uses the column density 
and magnetic field,
\begin{eqnarray}
\lambda = 2 \pi G^{1/2} { \Sigma \over  B}. 
\end{eqnarray}

Due to the anisotropic nature of the magnetic field, and in particular to the fact that the Lorentz force vanishes along the field lines, a magnetized cloud does not collapse spherically but 
typically get flatten along the field lines. It should be stressed however that as the collapse proceeds, the field lines get strongly pinched by the differential motions, forming an hour glass shape. This has two consequences. First, it creates a magnetic pressure force, 
$-\partial _r (B_z^2) / 8 \pi$, 
parallel to the equatorial plane and pointing outwards. This typically slows down the collapse when magnetic field is strong enough. Second, it creates another component of the magnetic force, which is oriented along the z-axis and is equal to $-\partial _z (B_r^2) / 8 \pi$.  However, for symmetry reasons, $B_r$ is usually vanishing in the equatorial plane. Therefore, this force is pointing toward the mid-plane and is compressing the gas. 
Compared to the hydrodynamical collapse, for which the singular isothermal, $\rho = c_s^2 / (2 \pi G r^2)$ is usually an acceptable approximation, particularly for low mass cores,  this creates close to the equatorial plane, a density enhancement which can be as high as a factor of several.  Because of its shape, this thin layer is usually called a pseudo-disk.  Exact solutions of this equilibrium, called the magnetized isothermal toroids, have been obtained by \citet{Li96}, while an approximation of the density enhancement and the dense layer thickness is presented in \citet{hf2008}. 
It should be stressed that like for the singular isothermal sphere, the resulting equilibrium is unstable to collapse and thus, unless the core is magnetically dominated, the pseudo-disk collapses. Indeed, the pseudo-disk, unlike a centrifugally supported disk, is not a structure which is at mechanical equilibrium. It is essentially a collapsing envelope that has been flattened by the magnetic field effect. 

It is worth stressing that the aligned configuration, i.e., magnetic field and rotation being initially parallel, which has generally been assumed in most early calculations, is somehow peculiar because in this configuration only the z-component of the field is not zero in the equatorial plane and therefore the magnetic pressure is squeezing the gas along the z-axis. If magnetic field and rotation are initially misaligned, the magnetic field lines get twisted by rotation and the radial field does not vanish anymore in the equatorial plane. This has as effect to somehow thicken the pseudo-disk
\citep{Hennebelle09,hirano2020}.

\subsection{Magnetic Braking}
\label{magbrak}
The magnetic braking acts through the magnetic torque $-r B_\phi {\bf B}$. Since it is a local quantity, it is not straightforward to anticipate its role, and it is enlightening to investigate some of its key properties thanks to simple cases, as proposed by \citet{Mouschovias80} 
and more recently by \citet{2012A&A...543A.128J} and \citet{hirano2020}. 

\subsubsection{Aligned rotator}
Let us first consider a cloud of mass $M$, density $\rho_c$, radius $R_c$, and half-height $Z$, surrounded by an external medium of density $\rho_{\rm ext}$. The cloud is initially in rotation at 
a speed, $\Omega$, and for simplicity the magnetic field ${\bf B}$ is considered uniform and parallel to the rotation axis. The magnetic braking timescale, $\tau_{\parallel}$, is estimated as the time needed for a torsional Alfv\'en wave to transfer the initial angular momentum of the cloud to the external medium \citep{mouschovias1979}:
\begin{equation}
\rho_{\rm ext}v_{\rm A,ext}\tau_{\parallel} \sim \rho_{c}Z. \label{eq:parallel}
\end{equation}
The Alfv\'en speed in the external medium is given by $v_{\rm A,ext} = B/\sqrt{4\pi\rho_{\rm ext}}$ , the mass of the core is $M \sim 2\pi\rho_cR_c^2Z$, while the magnetic flux is $\Phi_B \sim \pi R_c^2B$. This leads to 
\begin{equation}
\tau_{\parallel} \sim \left(\frac{\pi}{\rho_{\rm ext}}\right)^{1/2}\frac{M}{\Phi_B}. \label{eq:tau_classic}
\end{equation}
Therefore the magnetic braking timescale depends on the cloud environment through the density of the external medium, $\rho_{\rm ext}$. This emphasizes that magnetic braking represents a transfer 
of angular momentum between a faster rotating cloud and a slowly rotating surrounding medium. 

However, as discussed above, in a collapsing cloud, the magnetic field lines are strongly bent 
and therefore in Eq.~(\ref{eq:parallel}), one should take into account that as
the Alfv\'en waves propagate away from the clouds, the field lines are fanning out. 
Moreover, at equilibrium the field lines are likely in corotation, otherwise, 
the field toroidal component will be further growing. This leads to modify 
Eq.~(\ref{eq:parallel}) as
\begin{equation}
\Omega R_0^2\ \pi R_0^2\ \rho_{\rm ext}v_{\rm A,ext}\tau_{\parallel,fo} \sim \Omega R_c^2\ \pi R_c^2\ \rho_cZ,
\end{equation}
where $R_0$ is the typical distance between the field lines and the rotation axis, in the external medium.  We thus obtain the magnetic braking time for the case of fan-out, $\tau_{\parallel,fo}$, which is given by
\begin{equation}
\tau_{\parallel,fo} = \frac{\rho_c}{\rho_{\rm ext}}\frac{Z}{v_{\rm A,ext}}\left(\frac{R_c}{R_0}\right)^4. \label{eq:tau_alfo}
\end{equation}
It is similar to Eq.~(\ref{eq:parallel}), apart from the term $(R_c/R_0)^4$, whose origin is twofold. First, because the field lines fan-out, the volume of the external medium swept by the Alfv\'en waves is larger by $(R_0/R_c)^2$ than when the field lines are straight.  Second, because it is assumed that the field lines are in  co-rotation, the fluid elements which are attached to the field lines have a specific angular momentum that increases like $R_0^2$. This altogether leads to another factor $(R_0/R_c)^2$.

Combining Eq.~(\ref{eq:tau_alfo}) with the expressions of 
the mass, $M \sim 2\pi\rho_cR_c^2Z$, the magnetic flux of the core $\Phi_B \sim \pi R_0^2B_{\rm ext} = \pi R_c^2B_c$, and the expression for $v_{\rm A,ext}$, we obtain \citep[see, e.g.][]{Mouschovias85},
\begin{equation}
\tau_{\parallel,fo} = \left(\frac{\pi}{\rho_{\rm ext}}\right)^{1/2}\frac{M}{\Phi_B}\left(\frac{R_c}{R_0}\right)^2. \label{eq:tau_alfomf}
\end{equation}
Compared to the magnetic braking time with straight lines, the magnetic braking timescale is thus significantly reduced since in a collapsing core $R_c \ll R_0$ and $(R_c/R_0)^2$ is expected to be a small number. 

\subsubsection{Perpendicular rotator ($\alpha=90^{\circ}$)}

In the case of a perpendicular rotator, i.e., if magnetic field lines are initially perpendicular to the rotation axis,  the braking timescale corresponds to the time it takes for the Alfv\'en waves to reach $R_{\perp}$, the radius at which the angular momentum of the external medium is equal to the initial angular momentum of the cloud. The Alfv\'en waves propagate in the equatorial plane and sweep a cylinder of half-height $Z$ and radius $R_{\perp}$, thus
\begin{equation}
\rho_{\rm ext}(R_{\perp}^4 - R_c^4) \sim \rho_c R_c^4. \label{radius}
\end{equation}
Considering that the magnetic field is such that $B(r) \propto r^{-1}$, so that $v_{\rm A}(r) = v_{\rm A}(R_c) \times R_c/r$, the magnetic braking time is then
\begin{equation}
\tau_{\perp} = \int_{R_c}^{R_{\perp}}\frac{{\rm d}r}{v_{\rm A}(r)} \\ = \frac{1}{2}\frac{R_c}{v_{\rm A}(R_c)} \left[\left(1 + \frac{\rho_c}{\rho_{\rm ext}}\right)^{1/2} - 1 \right], \label{eq:tau_perpmf1}
\end{equation}
which leads to
\begin{equation}
\tau_{\perp} \sim 2\left(\frac{\pi}{\rho_c}\right)^{1/2}\frac{M}{\Phi_B}. \label{eq:tau_perpmf}
\end{equation}

Comparing the braking timescales between the aligned configuration with straight field lines and 
the perpendicular rotator,  Eq.~(\ref{eq:tau_classic}) and (\ref{eq:tau_perpmf}) gives 
\begin{equation}
\frac{\tau_{\parallel}}{\tau_{\perp}} = \frac{1}{2}\left(\frac{\rho_c}{\rho_{\rm ext}}\right)^{1/2}. \label{comp1}
\end{equation}
Since $\rho_c \gg \rho_{\rm ext}$,  magnetic braking is found to be more efficient in the perpendicular case. However, when taking into account the fact that field lines are fanning out, the 
term $(R_c/ R_0)^2$ must be taken into account.
As an illustrative example, let us assume that the density follows the singular isothermal sphere, 
i.e.  $\rho_{\rm ext}(R_0) \propto R_0^{-2}$, and  $\rho_c(R_c) \propto R_c^{-2}$. With Eqs.~(\ref{eq:tau_alfomf}) and (\ref{eq:tau_perpmf}), we obtain
\begin{equation}
\frac{\tau_{\parallel,fo}}{\tau_{\perp}} = \frac{R_c}{R_0}.
\end{equation}
Since $R_c/R_0 \ll 1$, the magnetic braking time is thus shorter in an aligned rotator than in a perpendicular one provided the field lines are fanning out.

Numerical simulations of misaligned collapsing cores have been performed by \citet{matsumoto2004}, \citet{2012A&A...543A.128J}, \citet{tsukamoto2018} and \citet{hirano2020}
where  analysis of the angular momentum distribution through the core were also performed.
The reported results have been in apparent contradiction. 
While \citet{matsumoto2004} and  \citet{tsukamoto2018}  concluded that the angular momentum in the core was higher in the aligned configuration than in the perpendicular one, 
\citet{2012A&A...543A.128J} and \citet{hirano2020} concluded the opposite. 
The origin of this apparent contradiction lays on the different timescale that are 
being probed in these studies. \citet{matsumoto2004} and  \citet{tsukamoto2018}  
analyse their simulations at a relatively early time, that is to say before any disk forms. 
They also analyse the angular momentum within very dense material. However, \citet{2012A&A...543A.128J} and \citet{hirano2020} investigate later times, after a disk has formed and looked at the 
angular momentum distribution through the core. The most likely explanation  is that 
at early times and for the densest material, the field lines have not been strongly stretched
and thus the braking time is shorter in the perpendicular configuration than in the parallel one. 
However, as time goes on, the material that falls in the central region, say the star/disk material, was initially located further away and thus the corresponding field lines are more and more bent leading to
a stronger braking.

\section{The formation of centrifugally supported disks}
In recent years, most of the efforts regarding the collapse of dense cores has been devoted to the study of disk formation. This has appeared to be a complex topic, which in particular relies on the assumptions made regarding the microphysics. 
Here we follow a progressive approach starting from the basics.

\subsection{What is the origin of angular momentum in star formation models?}
The very first question to be asked when discussing 
rotationally supported disks is obviously what is the 
value of angular momentum to be considered and where it comes 
from. Traditionally the working assumption has been 
that angular momentum was inherited from large scales and that 
it was conserved, at least to some degree, during the collapse. 
While this picture is widely accepted, it should be stressed 
that, in fact, it partly relies on the underlying assumption that the collapsing cloud is axisymmetric, in which case the 
distinction between radial and azimuthal velocity fields can be 
made rigorously. Moreover, observationally infall and rotational 
motions can be made easily. 
However, in the general case, clouds are not 
axisymmetric and this has important consequences. First, 
while angular momentum is still conserved with respect to the center of mass, the latter is usually not a relevant point because it is not the collapse center. Moreover, the star that forms is not attached to the center of mass, but its position moves as it evolves, creating small offsets between the global center of mass of the envelope+star system, and the star itself. Therefore, angular momentum is not a conserved quantity with respect to 
the star. From the inertial forces, a torque is actually operating.
This implies that angular momentum may not necessarily need 
to be inherited from the large scales. Indeed, \citet{2020A&A...635A.130V} 
investigated the collapse of a non-axisymmetric cloud which initially
had no motion and therefore no angular momentum. They show that 
in this configuration, disks would form as well.

In the rest of the section, we will not consider this 
scenario further, but it will be discussed later in this review.

\subsection{Disk formation in hydrodynamical models}
Let us consider a rotating and axisymmetric cloud without 
magnetic field. In such circumstance, angular momentum is a conserved quantity. 
Let  $M_*$ be the stellar mass and $j$, the fluid particle  specific angular momentum. 
The centrifugal radius, $r_d$, is such that centrifugal and gravitational forces 
compensate each other, leading to
\begin{eqnarray}
r_d = {j^2 \over GM_*},
\label{centrifug}
\end{eqnarray}
which clearly shows that disk formation is directly related to 
angular momentum distribution.

Let us consider a cloud with a density profile initially proportional to 
$1/r^2$ \citep[e.g.][]{Shu77} leading to a mass, $M(R_0)$, within  
radius $R_0$, such that $M(R_0) \propto R_0$. On the other-hand, the specific angular momentum 
is given by
$j(R_0) =  R_0^2 \Omega$. Therefore, we get $j(R_0) \propto M(R_0)^2$.  The centrifugal radius 
for a fluid particle initially located in $R_0$ can therefore be expressed as 
$r_d = j(R_0)^2 / (G M(R_0)) \propto M(R_0)^3$ \citep{Terebey84}.
On the other hand, considering a uniform density, $M(R_0) \propto R_0^3$ and $j \propto M(R_0)^{2/3}$ 
which gives $r_d \propto M(R_0)^{1/3}$. If for simplicity a constant accretion rate is assumed, 
 the disk is found to grow like $t^{3}$ when the density is initially in $r^{-2}$ 
 and to grow like $t^{1/3}$ when it is uniform.

From these two examples, it is seen that the angular momentum distribution sensitively determines  the disk radius. 
To get a quantitative estimate useful for reference, let 
$\rho_0$ be the cloud density, and $\Omega_0$ the angular rotation 
velocity.
For  a fluid particle initially located at radius $R_0$, 
the centrifugal radius is given by
\begin{eqnarray}
{r_\mathrm{d,hydro}\,\simeq\,{\Omega_0^2 R_0^4 \over 4 \pi / 3 \rho_0 R_0^3 G }\,=\,3 \beta R_0}  \\= 106 \, {\rm AU} \, {\beta  \over 0.02 }\, \left( {M \over 0.1 M _\odot}\right) ^{1/3}  \left( {\rho_0 \over 10^{-18} {\rm g \, cm}^{-3} }\right)^{-1/3},
\nonumber
\end{eqnarray}
where $\beta =R_0^3 \Omega_0^2/ 3GM$ is the ratio of rotational over gravitational energy.  
Note that observationally, cores have been initially inferred 
to have a typical $\beta \simeq 0.02$ \citep{Goodman93, Belloche13a}, which is the value used for reference.

\subsection{Disk formation in ideal MHD models}

The first collapse calculations, which have been performed 
with a magnetic field, assumed to be  parallel to the 
rotation axis
\citep[e.g.][]{allen2003,Galli06,Price07,hf2008,Mellon09}. 
A surprising conclusion has been that even with relatively 
modest magnetic fields, typically corresponding to $\mu$ as high as 5-10, 
the disk formation was nearly 
suppressed, a process which has been called catastrophic 
braking.
 The reason of this behavior can be understood by using simple orders of magnitude.

The  magnetic braking  and the rotation time, which are 
most important, are given by
\begin{eqnarray}
\label{eq_br}
\tau _{\rm br} &\simeq& { \rho v_\phi 4 \pi h \over B_z B_\phi}, \\
\tau _{\rm rot} &\simeq& { 2 \pi r \over v_ \phi}~,
\label{eq_rot}
\end{eqnarray}
where $h$ is the scale height of the disk.  If the rotation time is longer than the braking time,
a fluid particle may rotate a few times before it loses a significant amount of angular 
momentum. Since the magnetic torque is proportional to $B_\phi$, we need to estimate it, which 
can be accomplished using the Maxwell-Faraday equation. Essentially, $B_\phi$ is produced 
by the twisting of the poloidal magnetic field by the differential rotation. 
As long as ideal-MHD holds, it grows continuously with time, and we have
\begin{eqnarray}
B_\phi \simeq \tau _{\rm br} { B_z v_\phi \over h}.
\label{bphi}
\end{eqnarray}

This leads to 
\begin{eqnarray}
{\tau _{\rm br} \over \tau _{\rm rot} } &\simeq& \left( { M_* \over M_{\rm d} } \times { h \over r_d } \right)^{1/2} 
{ G^{1/2} \Sigma \over 2  B_z  } \\ && \simeq \left(  { M_* \over M_{\rm d} }
{ h \over r_d } \right)^{1/2}   { \lambda _{eff} \over 4 \pi   }, 
\nonumber
\label{eq_br3}
\end{eqnarray}
where we have assumed $v_\phi \simeq \sqrt{ G M_*  / r}$, $M_*$ being the mass of the central star and 
where $M_{\rm d} = \pi r_d^2 \rho h$ is the disk mass, $\Sigma=2 h \rho$, is the disk column density and $\lambda_{eff} =  2 \pi G^{1/2} \Sigma / B$ is the mass to flux ratio. 
This shows that modest magnetic intensities corresponding to values as high as few times $ 4 \pi$, the magnetic braking  timescale is shorter than the rotation time and therefore should prevent or severely limit the formation of centrifugally supported disk in good agreement with what has been inferred 
from the simulations.

\subsection{Disk formation with non-ideal MHD}

Since observational estimates of the typical values of $\mu$ in dense cores indicate values 
of a few \citep{Crutcher12}, a very efficient magnetic braking capable of suppressing 
disk formation for most observed cores, is clearly in tension with observations. 
As mentioned in Sect.~\ref{magbrak},  misalignment between magnetic field and rotation 
\citep{2012A&A...543A.128J,li2013,gray2018,hirano2020}
could help to alleviate the problem by reducing the efficiency 
of magnetic braking. Indeed, disk formation is reported in  all studies
which have considered misalignment 
 for values of $\mu$ on the order of several, though
the exact value of $\mu$ at which disk would form, varies between studies.
We stress however that when a disk forms, if ideal MHD applies, the toroidal magnetic field within the disk is 
continuously amplified and therefore the disk quickly becomes thick and inflated \citep{Hennebelle08b}. This 
emphasizes the need to consider non-ideal MHD. 

A very important property of ideal MHD is that the magnetic flux through a given fluid particle
is conserved. When non-ideal MHD processes are important, significant deviations from flux freezing can be 
produced, and this may qualitatively change  the impact of magnetic braking. Two main classes of 
process may lead to strong departures from flux freezing. First, if the coupling between the 
magnetic field and the neutrals is poor and second, if the flow is strongly turbulent. Let us stress that the latter corresponds
to a high magnetic Reynolds number, while the former would correspond to a low magnetic Reynolds number.

\subsubsection{The possible role of turbulence}

Turbulence violates the frozen-in condition of
ideal MHD flows because it drives reconnection and thus, the diffusion of the magnetic field
lines. Since
the original proposition made by \citet{lazarian1999}, this has been extensively studied and demonstrated numerically  \citep[e.g.][]{Kowal2009, Eyink2013, santos2021}.

The role that turbulence may have on disk formation has been carefully investigated by \citet{santos2012}, \citet{seifried2012} and \citet{joos2013}. 
In all these studies, it has been concluded that the efficiency
of magnetic braking is reduced by turbulence, though the proposed explanations were not identical.
\citet{santos2012} emphasized the role of turbulent reconnection which transports magnetic field 
outwards therefore leading to a reduced magnetic torque, while \citet{joos2013} stressed that the misalignment induced by turbulence may add up to the 
reconnection diffusion triggered by turbulence in reducing magnetic braking efficiency. \citet{seifried2012} proposed that it 
may be the coherence of the averaged magnetic torque that is reduced by the fluctuating magnetic field, again induced 
by  turbulence. 
While it is clear that reconnection diffusion is occuring in simulations which include 
turbulence \citep[see][for a discussion]{santos2013}, it remains difficult to estimate the 
respective role played by these three effects, which all concur to favor disk formation.

\subsubsection{Low ionisation and high resistivities}

One of the most obvious solutions to the so-called magnetic braking catastrophe is 
to be searched in high magnetic resistivities. This has been early 
proposed by \citet{Galli06} where Ohmic dissipation was envisioned to limit magnetic braking. 
A long series of increasingly realistic calculations has since been performed by several 
authors. A comprehensive description of the various 
calculations performed can be found in \citet{zhao2020}. 

For instance, \citet{Dapp10} show that when Ohmic dissipation is included in their calculation a small disk forms but not 
when they assume ideal MHD. This is because at small scale, the magnetic field is efficiently 
diffused out and therefore the braking time typically becomes longer than the rotation time. 
Several works have performed 
simulations that include ambipolar diffusion, \citep{Dapp12,tomida2013,tomida2015,masson2016,zhao2016,hennebelle2020} 
finding that centrifugally supported disks of few tens of AU 
always form. Note that on the contrary, \citet{wurster2016} and \citet{2019MNRAS.489.5326L} conclude that ambipolar diffusion is not sufficient to produce a disk.

Analytical arguments to predict the radius of the disks have been developed by \citet{hennebelle2016}. 
The effect of ambipolar diffusion, $\tau _{\rm diff}$, is determined by a characteristic timescale which describes how fast $B_\phi$ is diffused out of the disk
\begin{eqnarray}
\tau _{\rm diff} &\simeq &{ 4 \pi  h^2 \over c^2  \eta_{\rm AD}  } {  B_z^2 + B_\phi ^2 \over B_z^2} \simeq { 4 \pi  h^2 \over c^2  \eta_{\rm AD}  }~.
\label{tau1}
\end{eqnarray}
To get a stationary $B_\phi$ within the disk, an equilibrium 
between generation and diffusion must occur. 
Combining Eqs.~(\ref{bphi}) and~(\ref{tau1}), as well as  Eq.~(\ref{eq_br}) with Eq.~(\ref{eq_rot}) together with an estimate of the density 
at the edge of the disk, 
\begin{eqnarray}
\rho(r) = \delta { C_{\rm s}^2 \over 2 \pi G r^2} \left( 1 + {1 \over 2} \left( { v_\phi (r)\over C_{\rm s}} \right)^2 \right). 
\label{rho_exp}
\end{eqnarray}
where $\delta$ is a coefficient on the order of a few, the following expression is inferred
\begin{eqnarray}
 r_{ad}  \simeq  18 \, {\rm AU}  \times
\delta ^{2/9 } \left( {   \eta_\mathrm{AD} \over 7.16 \times 10^{18} \, {\rm cm^2~s^{-1}}   } \right)^{2/9} \\ \times
\left( {B_z \over 0.1\,{\rm G} } \right) ^{-4/9}   \left( { M_\mathrm{d} + M_* \over 0.1 M _\odot} \right) ^{1/3}.
\nonumber
\label{radius}
\end{eqnarray}
This expression has been compared with a broad set of MHD simulations \citep{hennebelle2016,hennebelle2020} and an agreement within a factor of about 2 has been inferred for the simulations in which the magnetic field is strong enough. 
It shows in particular that the disk size grows with the magnetic resistivity and also with the stellar mass. 

The influence of the Hall effect (see Eq.~\ref{Eq:inductB}) has been the subject of 
several studies. Let us first note a peculiarity of the Hall term. Flipping the sign of ${\bf B}$
in Eq.~(\ref{Eq:inductB}), we see that all terms, but the Hall one, change their sign. Physically, this 
is because the Hall term describes the generation of the magnetic field induced by the motion of the 
charged particles induced by the Lorentz force. This implies that two configurations, i.e., the aligned and anti-aligned cases (${\bf \Omega} . {\bf B}  > 0$ and $<$ 0 respectively) have to be distinguished. As a consequence, several teams have found 
that in the aligned configuration, the magnetic braking is enhanced 
and, on the contrary, it is reduced. 
This includes the work of 
\citet{braiding2012} in which analytical solutions are being used, 
\citet{tsukamoto2015}, \citet{wurster2016} which perform 
3D smooth particle hydrodynamical simulations and 
\citet{marchand2019} which perform adaptive mesh refinement
calculations. Due to these different magnetic braking
efficiencies,  these authors find that the disks
that form in the aligned case are smaller, than the disks
which form when the Hall effect is not accounted for, which are 
themselves smaller than the disks which form in the anti-aligned
configurations. This has led to the idea of the possible existence
of a bimodal population of disks.
\citet{leemar2021} have extended the study of 
\citet{hennebelle2016} to include the Hall effect, predicting 
disks of similar size to that inferred in the simulations.
\citet{tsukamoto2017} present calculations of misaligned configuration and conclude that the impact of anti-alignment persists,
even so the differences between the 45$^\circ$ and $135^\circ$
is significantly reduced compared to the differences between 
0$^\circ$ and $180^\circ$.

Importantly, \citet{zhao2020} and \citet{2021MNRAS.505.5142Z} have run 
2D simulations for longer period of time and conclude that the
external part of the disks formed in the anti-aligned configuration, 
tend to disappear, leading to a few tens of AU disks. This would 
suggest that the bimodal distribution could be a transient feature.

Finally, we stress that both the high resistivities and the reconnection diffusion induced by turbulence could 
contribute simultaneously or in different situations. For instance, if in some dense cores, the ionisation fraction 
is high, the resistivities could be lower and the reconnection diffusion possibly dominant. However, so far the 
studies which have considered in isolated dense cores, both high resistivities and turbulence, have concluded that the latter does not substantially
modify the formation of disk \citep[e.g.][]{hennebelle2020,wurster2020}.

\label{sec:MHDmodels}

\section{Tracing magnetic fields with photons around protostars or at core scales}

The current perspective on magnetic fields properties in star-forming cores has been established observationally by the analysis of polarized light resulting from the interaction of the B fields with the surrounding gas molecules and dust grains. 
Several observational techniques have been developed: we detail them concisely here below, then describe how these measurements are used to obtain quantitative constraints from observed objects, and how the physical processes they rely on have been implemented in radiative transfer models to produce synthetic observations.

\subsection{Observational signatures of magnetic fields in protostars}\label{obs-Bsignatures}

Most of the observational signatures of the presence of the magnetic field in the interstellar medium are directly related with polarization. However, there are other mechanisms that can also produce polarization and are not related with the presence of magnetic fields. Here on, we focus on the magnetic fields signatures expected in molecular clouds at core's scales, the possible caveats and other sources of polarization for each case:

\subsubsection{Zeeman effect} 

Molecular rotational lines split in submagnetic levels under the presence of a magnetic field. The frequency separation of these levels is proportional to the magnetic field strength and the magnetic dipole moment.  For most molecules, the magnetic dipole moment is very small, making the Zeeman splitting undetectable, except for the maser lines \citep[e.g.,][]{Alves12, Crutcher19}. However, molecules with an unpair electron have relatively large magnetic moments, yielding to significant Zeeman splitting. The Zeeman molecules that are abundant in molecular clouds are OH, CN, SO, CCH and CCS. The splitting can be measured through circular polarization, but in almost all cases not in total intensity. This is because the Zeeman splitting in frequency is typically only of the order of Hz $\mu$G$^{-1}$ \citep{Bel89, Bel98, Shinnaga00, Uchida01, Turner06, Cazzoli17}, much smaller than typical linewidths in molecular clouds for the expected magnetic field strength ($\lesssim$ few mG). The circular polarization is proportional to the magnetic field strength component along the line-of-sight, the intensity derivative and the specific Zeeman splitting for the observed line.  The molecules with hyperfine structure (CCH, CN) are more adequate to observe the Zeeman splitting because different hyperfine levels have different Zeeman splitting values.  This allows to separate the possible instrumental polarization, which does not distinguish between hyperfine lines, from the Zeeman effect. There have been several attempts to detect Zeeman effect at core scales,  mostly through CN rotational transitions, but there are only a handful of detections reported in the literature \citep{Crutcher99a, Crutcher99b, Levin01, Falgarone08, Maury12b, Pillai16, Nakamura19}, and none at disk scales \citep{Vlemmings19, Harrison21}.  

{\it Caveats:} This is the only method that can provide a direct estimation of the field strength along the line of sight. Nevertheless, in order to be able to derive the role of the magnetic fields at the core and disk scale, special care has to be made on where the emission of the Zeeman molecules arise. These are radical molecules, so they are chemically active and their abundances may change significantly depending on the environment conditions. Thus, in prestellar cores, CN appears to behave as other N-bearing molecules and thus survives at high densities \citep{Hily-Blant08}. On the contrary, CCH, CCS and SO appear to deplete toward the core's center \citep{Tafalla06, Padovani09, Juarez17, Seo19}. For massive cores, all these molecules but SO appear to be anticorrelated with the dust emission at scales of $\lesssim 0.05$~pc \citep{Dirienzo15, Paron21}. The distribution of these molecules in the planet-forming disk is also complex and, combined with the expected magnetic field configuration, makes the interpretation of Zeeman observations difficult \citep{Mazzei20}.

\subsubsection{Linear polarization of molecular lines} 

Rotational levels split into magnetic sublevels under the presence of magnetic fields. Transitions where the submagnetic level does not change will be linearly polarized, with the polarization angle parallel to the magnetic field. In the other cases, the polarization will be perpendicular to the magnetic field. In most cases, collisions do not differentiate among magnetic sublevels, so they populate the sublevels equally. That means that the net linear polarization is zero. However, an anisotropic radiation field will generate unbalance of the sublevels, giving a partially linearly polarized emission. This process is known as the  Goldreich-Kylafis (G-K) effect because it was initially developed by \citet{Goldreich81}. 
The level of linear polarization produced by the G-K effect depends on various parameters, such as the ratio of the collisional rate to the radiative decay rate and  the optical depth \citep{Goldreich81, Goldreich82, Deguchi84}.  

However, the polarization direction can be parallel or perpendicular to the magnetic field.  This means that the properties of the emission (optical depth, velocity, and density gradients) should be analyzed to solve this ambiguity. Once this is done, then this technique allows obtaining the magnetic field morphology projected in the plane of the sky.  Multi-transition observations of the linearly polarized lines can be used to derive the field strength \citep{Cortes05}.
The G-K has been detected in a small sample of cores \citep{Girart04, Lai03, Cortes05, Cortes08, Cortes21}, in few disks \citep{Stephens20, Teague21} and in some molecular outflows  \citep{Girart99, Ching16, Lee18}. Differential collisions due to ambipolar diffusion could also produce polarization in the ion molecules \citet{Lankhaar20b}.

{\it Caveats:} The properties of linearly polarized emission from a molecular line produced by the G-K can be altered if there is a foreground molecular component at the same velocity \citep{Hezareh13}. This is the  Anisotropic Resonant Scattering (ARS) effect,  which not only alters the properties of the linear polarization, but it leaks a fraction of this, generating circular polarization \citep{Houde13, Houde22}. There is some evidence that this may happen \citep{Chamma18}. Sensitive observations of all Stokes parameters are needed to correct for this effect and obtain the unaltered original signature of the G-K effect. In addition, the detection of this non-Zeeman circular polarization could be used to measure the magnetic field strength \citep{Houde13}. Molecular ions could also have collisionally-driven linear polarization produced by ambipolar drifts \citep{Lankhaar20b}. The resulting polarization has an angle perpendicular to the magnetic field direction. This effect could be distinguished from G-K by using optically thin molecular ion lines.  

\subsubsection{Dust polarization} 

Since the early 1950s we know that interstellar grains are partially aligned with the magnetic fields \citep{Davis51}. The major axis of the grains are aligned perpendicular to the magnetic field, yielding  to linear polarization in the dust emission with an angle perpendicular to the magnetic field projected in the plane of the sky. There have been some proposed mechanisms  that allow the grain angular momentum to align with the magnetic field  \citep[e.g. see review by][]{Lazarian15}. Radiative torques (RATs) are thought to be the most efficient way to help grains to align with the magnetic field. Some predictions of this theory have been confirmed observationally \citep{Alves14, Jones15}. However, recent results with ALMA shows that at core scales RATs are not enough to explain the observed dust polarization properties \citep{LeGouellec20}. The dust emission appears to be polarized at significant levels ($\gtrsim 1$~\%) in a significant high fraction of observed cores, and therefore it is the  most used technique in the millimeter through far-IR wavelengths to study the magnetic fields at core and disk scales
(e.g., \citealt{Girart09}, \citealt{Pattle17}, \citealt{Alves18}, \citealp{Sanhueza21}, and for a more complete references see \citealp{HullZhang19} and \citealp{Pattle22}).

{\it Caveats:} At disks scales, grain growth is so important that self-scattering produced by large grains appears to be the dominant polarization mechanism \citep[e.g.,][]{Kataoka15, Yang17, Kirchschlager20}, although other mechanisms have also been proposed to generate polarization at disk scales \citep{Tazaki17, Kataoka19, Brunngaber19}. 
Of special interest for environments containing large grains, if feasible, is the the alignment of these grains with magnetic fields in the Mie regime, when grains start to have a size similar to or larger than the wavelength of the incident light \citep{Guillet20}. At the core scales, in the densest part where the dust emission may be optically thick, especially at the shortest wavelengths, the polarization pattern may be altered or even reversed \citep{Liu16, Ko20}.

\subsubsection{Ion to neutral velocity drift}

The ionization fraction in molecular clouds is tiny, $\sim 10^{-6}$, but the ions and neutrals are well coupled. However, small kinematic differences are expected due to diffuse processes such ambipolar diffusion \citep{Houde00,Li-HB08,FalcetaG10}. This effect can be observed as differences between neutrals and ions, in their line width and in the velocity maps. HCO$^+$ and HCN lines with the same rotational level from their different isotopologues are ideal to test this because they have similar excitation conditions.  The line width differences have been observed in several cores \citep{Houde00, Lai03b}. In a magnetized turbulent medium the linewidth difference should change with the length scale, which has been observed at $\sim 0.1$~pc scales, allowing to measure the turbulent dissipation scale and the magnetic field strength \citep{Li-HB08, Hezareh10, Hezareh14}. Recent observations of NH$_3$ and N$_2$H$^+$, at $\lesssim 0.1$~pc scales, show that, contrarily of what is expected in AD, the ion linewidth is systematically broader than that of the neutral by 20\% \citep{Pineda21}.

{\it Caveats:} The expected velocity drift is very difficult to measure at small scales typical of inner envelopes, as the velocity offset is expected to be of the order of the best spectral resolution currently available from typical instruments and the velocity field is complex \citep{Yen18,Cabedo22}. The main limitation of this method is to ensure that the selected neutral and ion species trace the same gas, or are affected by optical depth effects \citep[e.g.,][]{Pratap97, Jorgensen04, Girart05, Zinchenko09}.

\subsubsection{Observational techniques to infer magnetic field properties}

Here, we describe the main techniques that allow to infer the magnetic field properties from observations, independently of the use of the radiative transfer tools that are described in the next subsection. The most popular technique is the Davis-Chandrasekhar-Fermi (CDF) equation \citep{Davis51, Chandrasekhar53} which allows estimating the magnetic field strength in the scenario of small perturbation due to the anisotropic motions (such as turbulence). The equation relates the magnetic field in the plane-of-the sky, $B_{\rm pos}$ with the gas density, $\rho$, the non-thermal velocity dispersion along the line-of-sight, $\delta {\rm v}$ and the dispersion of the polarization angles of in the plane of the sky, $\delta {\rm \theta}$:  $B_{\rm pos} = \sqrt{4 \pi \rho} \, \delta {\rm v}/\delta {\rm \theta}$ \citep{Chandrasekhar53}. There have been several works to account for the limitations of this technique, such as line-of-sight smearing or cases with large dispersion, that in general leads to an overestimation of the field strength \citep{Ostriker01,Heitsch01,FalcetaG08, Cho16, Skalidis21, LiuJ21}.
A more sophisticated method is the use of the structure function, which allows separating the turbulent component from the smooth variations of the polarization angles to the large scale field \citep{Hildebrand09, Houde09, Houde11, Houde16}. 
There are other (semi) analytical expressions that allow to evaluate the relevance of magnetic fields for the specific hour glass configuration \citep{Ewertowski13, Myers18}.

\subsection{Polarized radiative transfer tools for protostellar environments}
In this section, we describe how physical processes responsible for producing polarized light as a result of the interaction of magnetic fields with protostellar material are implemented in radiative transfer codes and coupled to the numerical models for star and disk formation. We focus on two codes which are widely used by the community interested in star formation: POLARIS for dust grain alignment \citep{Reissl16}, and PORTAL for spectral lines \citep{Lankhaar20}.
While a complete description of the polarization-inducing processes go beyond the scope of this review, we briefly describe them below.

\subsubsection{Radiative transfer of emission from magnetically aligned grains}

\begin{figure*}[!h]
\begin{center}
\includegraphics[width=0.8\linewidth]{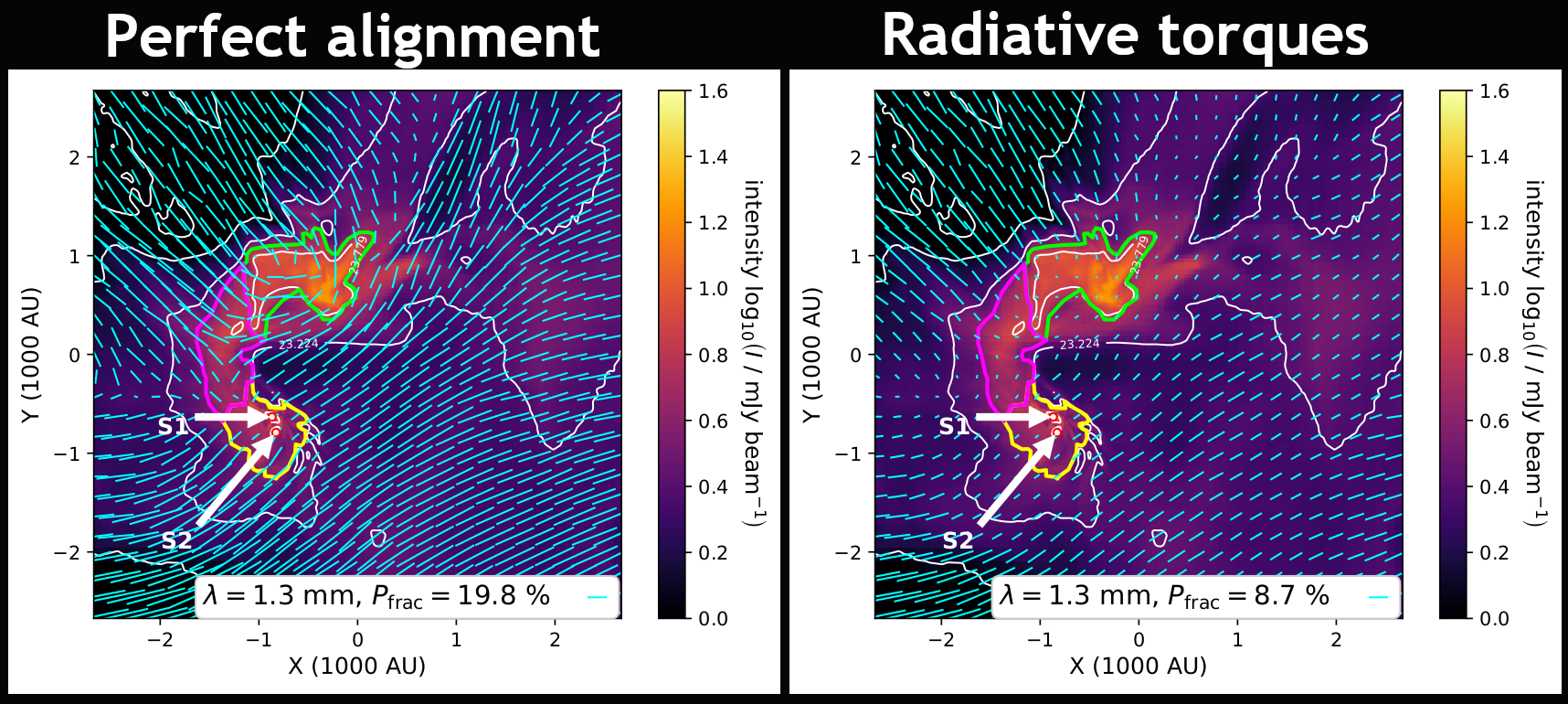}
\end{center}
\caption{Synthetic maps of the $1.3\ \mathrm{mm}$ dust continuum emission. Overlaid are the polarization vectors rotated by $90^\circ$, obtained with POLARIS either considering perfect alignment of the dust grains (right) or RAT grain alignment (left), from \citet{kuffmeier2020}.}
\label{fig:synthobs-kuffmeier}
\end{figure*}

The three-dimensional continuum radiative transfer RT code POLARIS \citep{Reissl16} uses 3D models of astrophysical structures to produce synthetic maps of the polarized dust continuum emission, allowing several flavours of alignment mechanisms for the dust grains. The photon propagation within the 3D model is implemented following a Monte-Carlo photon transfer scheme. The interaction of the incoming radiation with dust grains is determined in each cell, and depends on mostly of the cross-section of extinction, the dust density. Depending on the local physical conditions such as the dust grain albedo, POLARIS either computes the scattering, or the absorption with instantaneous re-emission, then a new wavelength is calculated and the dust thermal energy of the cell is adjusted. In this computation of photon propagation and dust heating mode, POLARIS assumes that dust grains are spherical: this step thus allows to derive the dust temperature in each cell with a limited computational effort. 
The alignment probability of dust grains in each cell is calculated, accounting for the imperfect internal alignment and the imperfect alignment between the dust grain’s angular momentum and magnetic field. Then, the anisotropy of the radiation field at any given wavelength $\lambda$ is calculated in each cell which allows determining the alignment of the different dust grains due to radiative torques (RATs, see previous section). Since a distribution of dust grain sizes is considered (usually following a classical MRN, after \citealt{Mathis77}, distribution consisting of power-laws of separate populations of bare spherical silicate and graphite grains), the RAT alignment in given irradiation and B-field conditions depends also on the effective grain radius $a_{align}$. Dust grains larger than $a_{align}$ are considered aligned (with a fraction depending on the high-J attractor point $f_{high-J}$, which can be manually set to any value between 0 and 1), and contribute to the polarization of the dust thermal emission. Several studies have used the POLARIS code to predict and confront observations of polarized dust emission, from molecular cloud conditions \citep{LiuJ21,Reissl21} down to disk conditions \citep{Brunngaber21}. Note that a few studies have used radiative transfer tools developed in a recent past, such as Dustpol \citep{Padovani12,2017ApJ...834..201L} but here we focus on the most recent results, and hence on POLARIS because it is the most widely used RT code for dust polarization nowadays. An example of synthetic observations of the polarized dust emission, assuming different alignment conditions for the dust from the same numerical model of a collapsing core, is shown in Figure \ref{fig:synthobs-kuffmeier}. Details and results of the works specifically dedicated to protostars are described in the next Section \ref{ssec:synthobs} below.

\subsubsection{Radiative transfer of emission from magnetically-sensitive molecular lines}

The POLARIS code can also be used to simulate the polarization of spectral lines due to the Zeeman effect, thanks to the ZRAD extension \citep{Brauer17}. This is based on the line RT algorithm Mol3D \citep{Ober15}, and makes use of atomic and molecular parameters such as the energy levels and transitions taken from the Leiden Atomic and Molecular Database (LAMDA), the Landé factors of the involved energy levels, the line strengths of the allowed transitions between Zeeman sublevels, and emitting radius of the molecular species. It has been used to test the robustness of Zeeman measurements to infer magnetic field strengths in molecular clouds, showing that while the gas density affects little the uncertainty of the measurements, strong variations in the LOS component of the gas velocity and of the magnetic field strength significantly impact the precision of the method \citep{Brauer17}.

PORTAL is an adaptive three-dimensional polarized line radiative transfer model that considers the local anisotropy of the radiation field as the only alignment mechanism of molecular or atomic species to calculate the aligned molecular or atomic states. It allows simulating the polarization of photons produced from line emission through a magnetic field of arbitrary morphology. While it can be run in stand-alone mode, PORTAL can also process the outputs of 3D radiative transfer codes on regular grids. It has been used, coupled to the parallelized non-LTE 3D line radiative transfer code Line Emission Modelling Engine (LIME, \citealt{Brinch10}), to predict  spectral line polarization of different molecular transitions arising from protoplanetary disks, at the ALMA wavelengths \citep{Lankhaar22}.

\subsection{Synthetic observations of B-fields from protostellar models: methods}
\label{ssec:synthobs}

The question of the physical processes causing the polarization of photons from star-forming structures, both from the dust and from molecular species, has been a long-standing one. Some simple analytical models assuming a quasi-perfect alignment of dust grains with magnetic field lines, for example, have been proposed in the past, and sometimes even successfully reproduced some observations \citep{Padovani12}. 
The development of more detailed physical models of grain alignment have allowed more predictive studies of the polarized dust emission arising from magnetically-aligned grains, and quantitative confrontation to observations performed towards protostellar envelopes.

The combination of polarized radiative transfer, as described above, to state-of-the-art numerical models of protostellar formation can be used to compare model predictions to actual observations.
The general process is quite standard. First, one selects the outputs corresponding the best to the properties of the observed object to be compared to. Usually, a numerical simulation output can be extracted as a datacube containing the gas density, the gas pressure (or equivalently the gas temperature), the three components of the velocity field, as well as the three components of the magnetic field. These gas properties are extracted and put in a suitable format to be post-processed with a radiative transfer code, such as \textsc{Polaris}. The central radiation source is modelled as a blackbody of chosen luminosity that allows to reproduce the observed bolometric luminosity of the studied object.
Note that, to facilitate the propagation of photons and avoid missed photon packages from the central source due to long computational times, many authors choose to artificially empty a small central sphere of radius a few au around the sink particle, as in \cite{Valdivia19}. The radiative transfer can also include an external isotropic radiation field of strength that illuminates the external layers of the core. 
The dust properties are normally taken from tabulated values and in most studies assumed uniform throughout the model, with a gas-to-dust ratio of $100$, and a typical composition and size distribution reproducing observations of dust in the ISM \citep{Mathis77,Hildebrand95}.
The radiative transfer calculation is done via Monte Carlo methods or ray tracing methods. For example, the POLARIS code computes first the propagation of photons alongside the dust temperature via a MCMC analysis, and then solves the grain alignment equations, using the density, radiation field, dust grain properties and temperature, in each cell of the grid. Grain alignment can be assumed to be perfect, or to depend on local conditions, as, e.g., when assuming RAT alignment \citep{Hoang14}. The outputs are maps of the Stokes I, Q and U from the dust thermal emission, which can be further convoluted by instrumental effects (e.g., the simobs task in CASA to simulate ALMA datasets) and compared to observations.

\label{sec:Btechniques-synthobs}

\section{Observed protostellar properties}
\label{sec:protostars-obs}
Measuring the properties of the gas, the dust, and the magnetic field in embedded protostars is cornerstone to inform the magnetized models of star formation, test their predictions, but also understand the role of non-ideal effects in disk formation and evolution. We recall here below some key observational results which relate directly to these questions.

\subsection{Dust Grains}
\label{ssec:dust-ion-obs}

Because they are the seeds from which planet formation processes are triggered in circumstellar disks, and because they are also key agents in the efficiency of coupling  the magnetic field to the gas, characterizing dust grains is a cornerstone in building a comprehensive scenario of star formation.

In the far-IR and longer wavelengths, the opacity of astrophysical dust scales with wavelength as $\kappa_{\nu}\propto \nu^{\beta}$, where $\beta$ is the emissivity spectral index of dust emission. $\beta$ reflects how emissive dust grains are, and is therefore,  commonly used to characterize dust grains in astrophysical structures.
It is measured by comparing the relative intensity of dust thermal emission when observed at different wavelengths: indeed, if the dust emission is optically thin, the ratio of flux densities at different wavelengths only depends on $\beta$. In the diffuse interstellar medium, Planck studies have shown the dust has $\beta$ $\sim$ 1.6 (Planck collab. 2014, Juvela et al. 2015).
Observational works show that significant variations in $\beta$ is found at protostellar core scales, with millimetre-wavelength $\beta$ values much lower than this typical value, however \citep{Martin12,Chen16,Sadavoy16,Bracco17,Vandeputte20}. 

\begin{figure*}[!h]
\begin{center}
\includegraphics[width=0.4\linewidth]{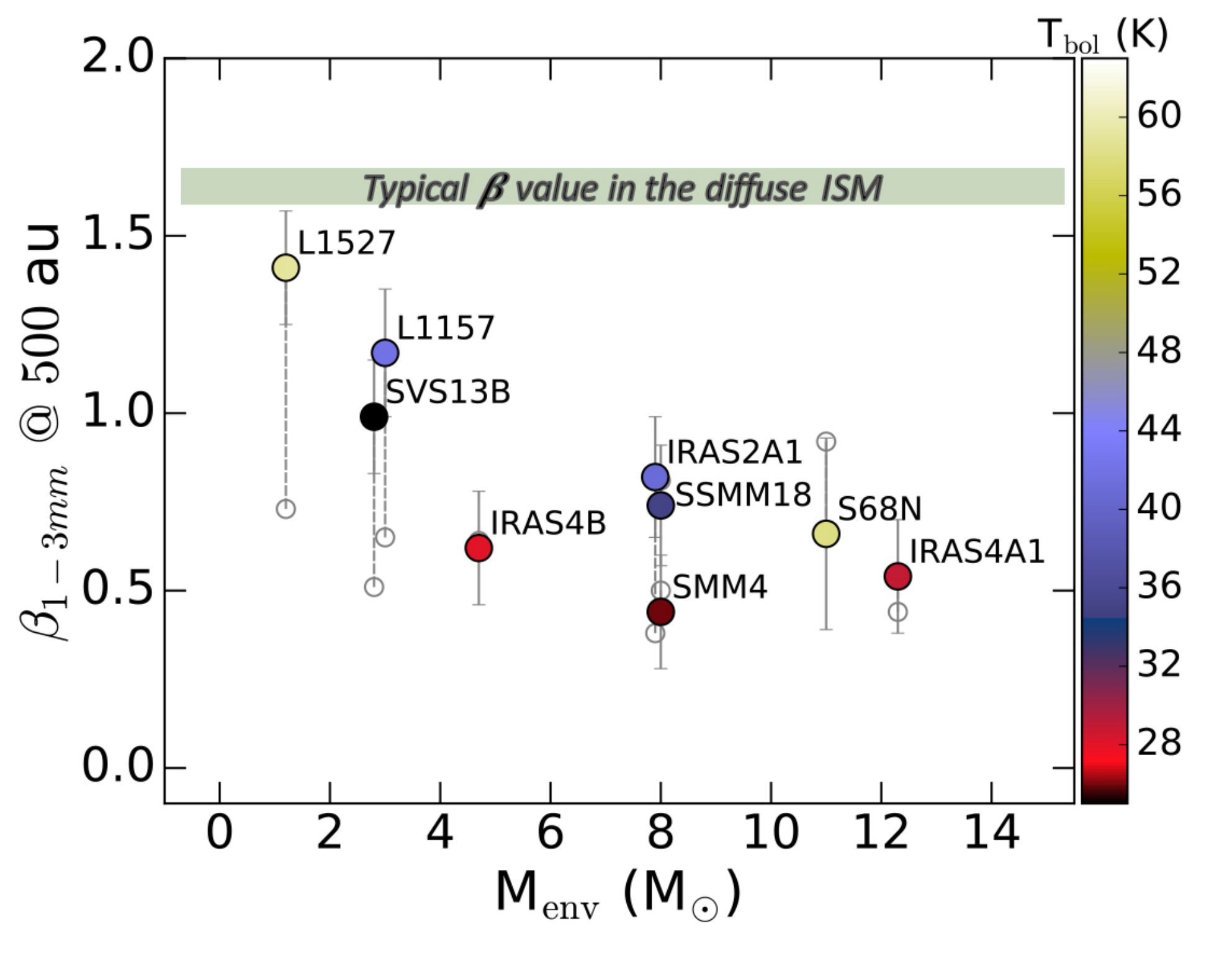}
\hspace{0.1cm}
\includegraphics*[width=0.55\linewidth]{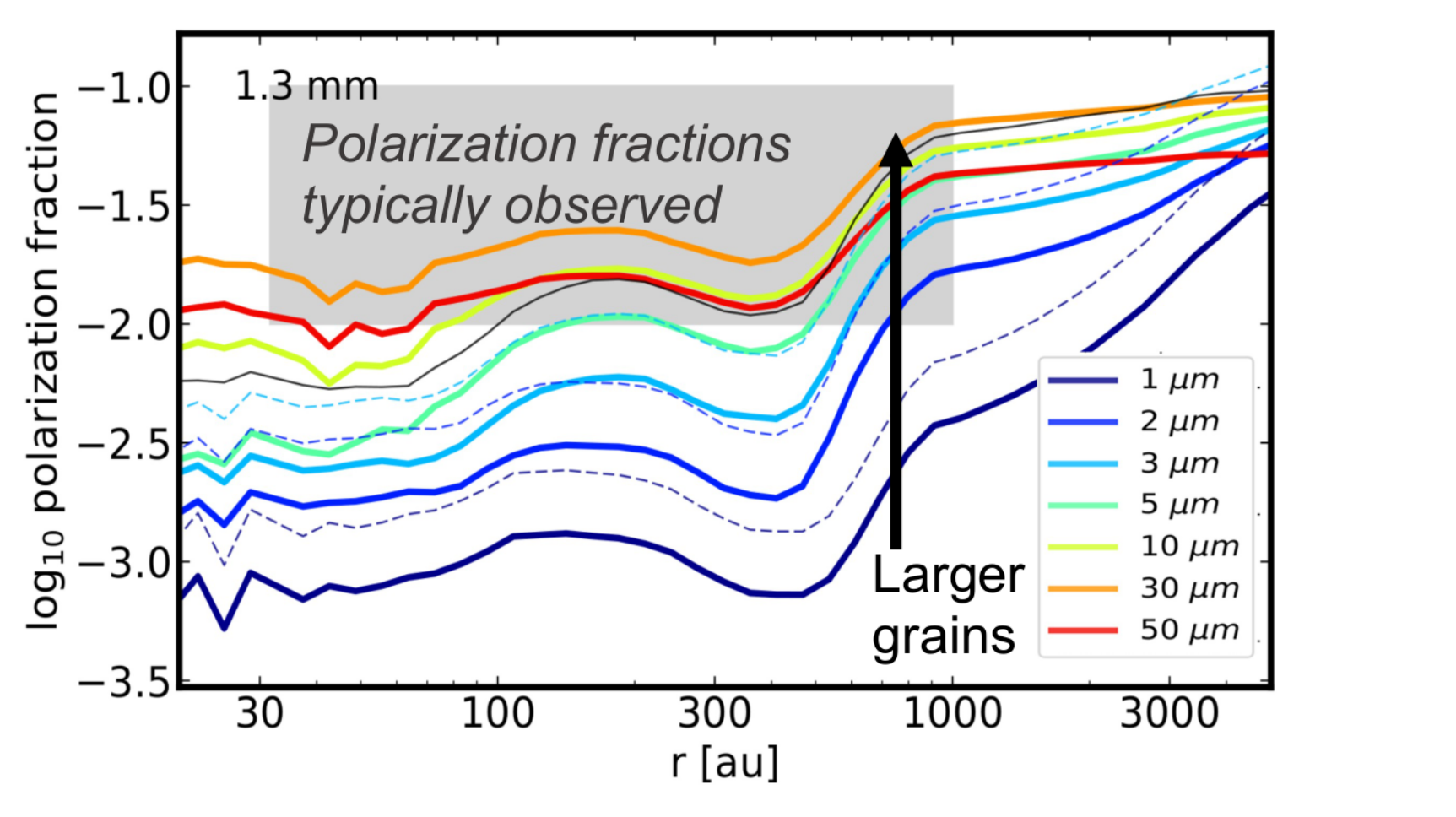}
\end{center}
\caption{Left: Observed millimeter dust emissivity index in a sample of young protostars from \citet{Galametz19}. All protostellar envelopes probed at radii $\sim 500$ au show lower values than the progenitor diffuse ISM. Right: Radial profile of the polarization fraction from the $1.3~\mathrm{mm}$ dust emission predicted from RAT alignment of dust grains in a typical low-mass protostellar envelope \citep{Valdivia19}. Large ($>10 \,\mu$m) dust grains are required to match the observations of the polarized emission (grey area).}
\label{fig:galametz_valdivia_dust}
\end{figure*}

While studies dedicated to resolving the dust properties at small radii in embedded protostars \citep{Jorgensen07a, Kwon09, Chiang12, 2014A&A...567A..32M} started before the advent of large millimetre interferometers such as NOEMA and ALMA, the development of these two observatories has allowed a significant leap forward in this area of research. 
\citet{Li17}, \citet{Galametz19}, \citet{2020A&A...640A..19T} and \citet{Bouvier21} have measured the dust emissivity spectral index in relatively large samples of embedded protostars, finding $\beta$ values ranging between 0 and 2, with a majority of the protostellar dust exhibiting $\beta < 1$ at envelope radii $\sim 500$ au. 
Figure \ref{fig:galametz_valdivia_dust} shows in its left panel the observations of millimeter dust emissivity indices measured by \citet{Galametz19} in the CALYPSO survey, finding that most protostars show significantly shallow dust emissivities at envelope radii $\sim 100-500$ au, with $\beta < 1$. 
\citet{Galametz19} did an analysis of their interferometric data in the visibility space, allowing to also distinguish a radial evolution of the dust emissivity, with many objects showing a decrease of $\beta$ with decreasing envelope radius. These observations also confirm it is the dust pertaining to the innermost envelope, and not to the unresolved disks at smaller scales, which is different from the dust typically observed in the diffuse ISM. 
In Class I protostars, less observations sensitive to the inner envelope regions are available: despite the poor statistics they also tend to show low dust emissivity in a majority of objects, with sometimes extremely low values ($\beta<0.5$) at disk scales \citep[see e.g.][]{Harsono18,Nakatani20,Zhang21}. 

Other effects affecting the millimetre flux could be responsible for the low dust opacities found, such as anomalous emission from spinning dust grains \citep{PlanckXX11}, contamination by non-thermal emission or the dust self-scattering \citep{Liu19}.
At the typical envelope scales, however, both the effects of optical depth and non-thermal emission are expected to be negligible at millimetre wavelengths \citep{2020A&A...640A..19T}. 

Another possible explanation for the observed difference in the dust emissivity $\beta$ could stem from different dust grain composition compared to that of the diffuse ISM. For example, it has been suggested that grains with higher ratio of carbonaceous to silicate  would exhibit lower $\beta$ \citep{Jones17, Ysard19, Zelco20}. Similarly, grains that are less compact and more fluffy could also be associated to low emissivity \citep{Brunngaber21,Kohler08}. These explanations are not favoured as the unique cause for the observed low emissivities, however, because of the low values found by observations (around 0) can not be matched by current dust models, even with significant changes of composition and compactness. Indeed, laboratory studies probing different interstellar dust analogues suggest that dust emissivities $<$1 at millimetre wavelengths can only be produced by grains larger than 100 \mic\ \citep{Demyk17a,Demyk17b,Ysard19} in the typical physical conditions reigning in protostellar envelopes.

Another independent thread of evidence for the presence of relatively large grains in Class 0 envelopes and disks stems from polarimetric observations \citep{LeGouellec19,Lee21} and their comparison to synthetic observations \citep{Valdivia19,LeGouellec20}, performed as described in \ref{ssec:synthobs}. 
They show that relatively large grains are required to reproduce the observed level of polarization fractions (see an example in the right panel of Figure \ref{fig:galametz_valdivia_dust}). Indeed, if dust grains align following the radiative torques (RATs) theory, the radiation field in deeply embedded protostellar environments is not prone to align the small grains where polarization of the dust mm emission is routinely detected.  Only synthetic observations performed with dust grain size distributions including grains up to $a_\mathrm{max} \sim 20 ~\mathrm{\mu m}$ produce polarized dust emission at levels similar to those currently observed in solar-type protostars, at (sub-)millimetre wavelengths.

\subsection{Fraction of ionized gas}
\label{ssec:dust-ion-obs}

Characterizing  the ionization of the gas in the dense envelope material is critical to set constraints on the coupling of the magnetic field with the infalling-rotating envelope gas, and the role of diffusive processes, such as ambipolar diffusion or reconnection diffusion, to counteract the outward transport of angular momentum from the infalling-rotating envelope due to B-fields. 

Cosmic rays (CRs), mostly relativistic protons, are the dominant source of ionization in relatively dense molecular gas where ultraviolet radiation cannot penetrate \citep{2015ARA&A..53..199G}. Protostars are deeply embedded sources, where the gas ionization fraction can only be inferred using indirect chemical signatures. A handful of measurements were obtained in Class 0 objects at core scales (typical densities $n_{\rm H_{2}} \sim 10^{4}$ cm$^{-3}$): they suggest typical cosmic ray ionization rates $\chi_e \sim 10^{-17}-10^{-15} \, \rm{s}^{-1}$ with large uncertainties \citep{2014ApJ...790L...1C,2014A&A...565A..64P,2017A&A...608A..82F,2018ApJ...859..136F}.
In Class I protostars, only a few studies have been carried out, using the abundance of HCO$^+$ to estimate gas ionization from cosmic rays in two young disks around $\xi \sim 10^{-17} s^{-1}$, and suggest it is lower in the disk compared to the inner envelope \citep{2021A&A...646A..72H,Vanthoff22}. Such low values may suggest it is unlikely that the accretion through the disk to the central star could be driven by magneto-rotational instabilities.
In the Class 0 protostar B335, \citet{Cabedo22} used deuteration detected in molecular line emission maps to characterize the ionization of the gas at envelope radii $\lesssim$ 500 au (typical densities $n_{\rm H_{2}} \sim 10^{6}$ cm$^{-3}$), and found large cosmic ray ionization rate $\zeta$ between $10^{-16}$ and $10^{-14}$~s$^{-1}$.
Such high values seem inconsistent with the fiducial value if the interstellar cosmic rays flux is responsible for the gas ionization, as it should be efficiently attenuated while penetrating into the dense cores \citep{2018A&A...614A.111P}.
The observations of \citet{Cabedo22} also show the CRs ionization rate is increasing at small envelope radii, toward the central protostellar embryo. Several theoretical works have investigated the role of shocks at the protostellar surface and magnetic mirroring within the jets as efficient forges to accelerate locally low-energy cosmic rays, and their role in increasing the ionization rate of the shielded protostellar material \citep{2015A&A...582L..13P,2018ApJ...863..188S,2021A&A...649A.149P,2021ApJ...915...43F}. 
In B335, it seems that local acceleration of CRs, and not the penetration of interstellar CRs, may be responsible for the gas ionization at small envelope radii. This would imply that, in the inner envelope, the collapse transitions from non-ideal to a quasi-ideal MHD once the central protostar starts ionizing its surrounding gas, and very efficient magnetic braking of the rotating-infalling protostellar gas might then take place.

\subsection{Gas kinematics: angular momentum and mass infall rates}
\label{ssec:kin-obs}

\begin{figure*}[!h]
\begin{center}
\includegraphics[width=0.7\linewidth]{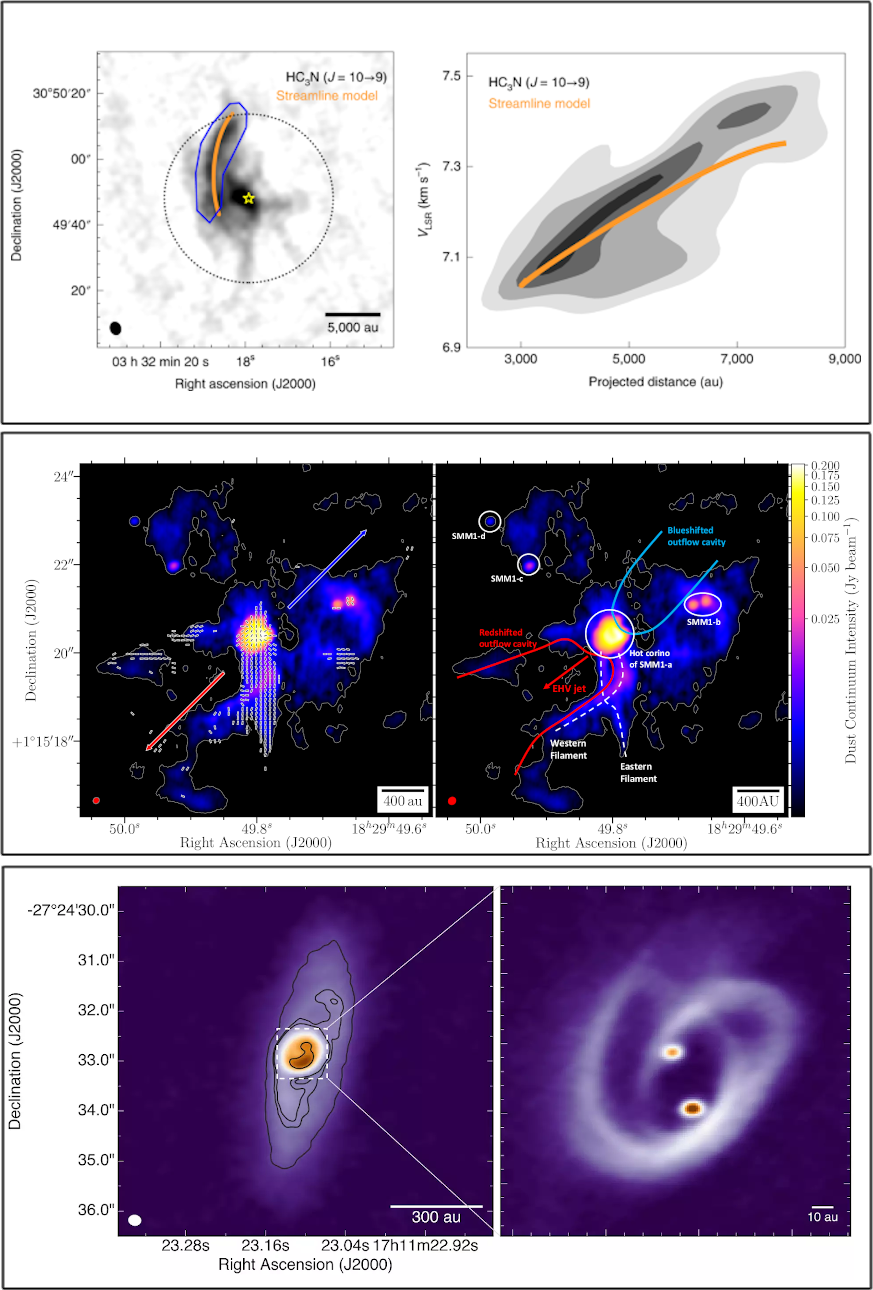}
\end{center}
\caption{Streamer structures seen at different scales and towards different embedded protostars. Top: IRAS 03292+3039 in Perseus \citep{2020NatAs...4.1158P}. Middle: Serp-SMM1 in Serpens \citep{LeGouellec19}. Bottom: [BHB2007] 11 in the Pipe \citep{2019Sci...366...90A}.}
\label{fig:streamers}
\end{figure*}

The dynamical role of the magnetic field can also be assessed by careful examination of the distribution of angular momentum associated to protostellar gas, and measurement of the mass transported inwards, from envelope to disk scales \citep{2019ApJ...882..103P,Galametz20, 2021ApJ...916...97Y}. 
This is not simple from an observational point of view, as the velocity field of the molecular gas is much less organized than expected from the simple monotoneous collapse of axisymmetric rotating cores in many protostars. Recent studies have revealed envelopes with complete reversal of the velocity fields, or containing multiple velocity components \citep{2020A&A...637A..92G,2017ApJ...849...89M} at radii $r<5000$ au. Figure \ref{fig:streamers} shows a few recent examples of the detection of well-developed asymmetric features from envelopes to disk scales, which may trace preferred pathways funneling the accretion, such as streamers  \citep{2020NatAs...4.1158P,Chen21} or supersonic infall along outflow cavity walls \citep{2021arXiv210701986C}.

Mass infall rates from large to small scales are difficult to estimate, as the complex kinematics of envelopes surrounding embedded protostars produce convoluted signatures whose degeneracies can only be lifted by a joint analysis of a variety of tracers probing widely varying density conditions and spatial scales. 
Double-peaked profiles in spectral lines from molecular species (due to self-absorption) with a brighter blueshifted peak are sometimes used as a signpost of probable infall motions \citep[e.g.,][]{Choi95, Mardones97, 2015ApJ...814...22E, 2015MNRAS.450.1926H, 2021ApJ...922..144Y}.
These profiles are used to derive infall rates, $\dot{M}_{\rm inf}$, following the expression: 
$\dot{M}_{\rm inf} = (\Omega/2) V_{\rm inf} \, (R_{\rm c}) \, \mu \,m_{\rm H} \,  \langle n_{\rm H_2}  \rangle \, R^2_{\rm c}$, 
where $\Omega$ is the solid angle subtended by the core center over which infall occurs, $R_{\rm c}$ is the core's radius, $V_{\rm inf} \, (R_{\rm c})$ is the infall velocity at this radius and $\langle n_{\rm H_2}   \rangle$ is the core's average volume density \citep[e.g.][]{Beltran22}. Surprisingly, few of these line profiles are observed towards embedded protostars at core scales \citep{Mottram13,Mottram17}.
Interferometric observations resolving the inner envelopes also revealed only a handful of them, for which modeling suggests typical mass infall rates $\sim 10^{-5}$ $\rm{M}_\odot$\ per year at $\sim500$ au scales in Class 0 protostars \citep{2012A&A...544L...7P,2015ApJ...814...22E,2019ApJ...885...98S}. Class I protostars in Ophiuchus have been surveyed with ALMA, exhibiting typical mass accretion rates a few $10^{-7}$ $\rm{M}_\odot$\ per year \citep{2019A&A...626A..71A}. 
In the well-characterized Class I protostar TMC1-A, the observed mass infall rate is measured to be larger than the value towards Ophiuchus Class I, estimated around $\sim 10^{-6}$ $\rm{M}_\odot$\ per year. This protostar also shows a decelerating infall velocity towards small radii, and an infall velocity $\sim 0.3$ times smaller than the free-fall velocity yielded by the dynamical mass of the protostar \citep{2015ApJ...812...27A}. 
The accretion rates derived in high mass star forming cores are significantly larger, typically between $\sim 10^{-4}$ to $10^{-2}$ $\rm{M}_\odot$\ per year at scales of few hundred to few thousands of au \citep[e.g.,][]{Beltran06, Qiu11, Wyrowski12, 2013ApJ...776...29L, 2020ApJ...904..181L}. In the case of the G31.41+.31, high angular resolution multi-line inverse P-Cygni measurements indicate infall acceleration at decreasing radius \citep{Beltran22}.

A key figure to assess the magnitude of rotational energy contained in protostellar cores, and compare these rotational motions to model predictions, is the specific angular momentum of the gas $j_{\rm{spe}}=r \times v_{\phi}$. 
High angular resolution observations of molecular line emission in Class 0 protostars have revealed that the gas contained in envelopes has a specific angular momentum $j_{\rm{spe}} \sim 10^{20}$ $\mathrm{cm}^2\,{{\rm{s}}}^{-1}\,$ at 1000 au \citep{2015ApJ...799..193Y,Tatematsu16,2019ApJ...882..103P,2020A&A...637A..92G,2021arXiv211209848H,Hsieh21}. 
In the literature, it is common to find that the conservation of angular momentum in an infalling rotating envelope should produce a radial profile of rotational velocity $v_{\phi} \propto r^{-1}$, because $j_{\rm{spe}}=r \times v_{\phi}$ should be constant. While if the initial rotation profile at the beginning of collapse follows a $v_{\phi} \propto r^{-1}$ relationship, this is true, it is not in most cases. Indeed, an infalling envelope initially in solid body rotation ($v_r \propto r^{-1/2}$ and  $v_{\phi} \propto r$) should exhibit $j_{\rm{spe}} \propto r^{2}$. If gas particles conserve their specific angular momentum during the collapse, the slope of the radial profile $j_{\rm{spe}}(r)$ should be conserved with time as they move inwards, and protostellar envelopes should thus be observed with $j_{\rm{spe}} \propto r^{2}$ down to the disk sizes. However, this is not what is observed.
At envelope radii between 1000 and 10000 au, $j_{\rm{spe}}$ scales as a power-law with radius, $j_{\rm{spe}} \propto r^{n}$ with $n\sim 1.6-1.8$. Interestingly, as pointed out by \citet{2020A&A...637A..92G}, this scaling is closer to the trend expected from the dissipation cascade of Kolmogorov-like turbulence than from solid-body rotation. This may suggest that large-scale ISM turbulence or turbulence locally driven around/by protostars is actually responsible for the observed angular momentum in outer envelopes.
Moreover, resolved observations have identified a break in the specific angular momentum radial profiles, from the steep outer profile $j_{\rm{spe}}(r) \propto r^{1.6}$ at envelope radii $r>1600$ au, to a quasi flat $j_{\rm{spe}}(r) \sim \rm{cte}$ inner profile at radii $<1600$ au \citep{2020A&A...637A..92G}. If angular momentum was conserved during the collapse, it suggests that the initial angular momentum profile in these cores was, also, mostly flat and would rule out that cores are build with solid body rotation as initial conditions.

Taken altogether, these observed features suggest that a complex interplay of magnetic fields, turbulence, and gravity are responsible for organizing the collapse along preferential directions, and that the angular momentum contained at small scales does not seem closely related to the gas flows observed at larger envelope radii.

\subsection{Magnetic fields at core's scales}
\label{ssec:B-obs}

High angular resolution observations at (sub-)mm wavelengths show that the thermal dust emission probing the envelope-to-disk scales ($\sim$ 50-10,000 au) is polarized at a few percent level. This has been observed both in low and high mass star forming regions \citep[see e.g.\,][]{Matthews09, Zhang14, Galametz18, HullZhang19, Beltran19, Sanhueza21, Eswaraiah21}. 
Considering that the polarized flux is a quantity that is prone to cancellation if the polarization angle is highly disorganized along the line of sight, observations of rather large polarization fractions suggests the magnetic field lines underlying the alignment of the protostellar dust remain at least partly organized inside star-forming cores \citep{LeGouellec20}. 
Indeed, these observations have already yielded to a large statistical sample to look for morphological trends at core scales. This is even taking into account that in many (mostly pre-ALMA) observations the detected signal was not extended enough to infer a clear, well resolved, morphology.
In many dense cores, the magnetic lines are often quite organized. Indeed, the expected hourglass configuration expected in the collapse of cores threaded by a uniform magnetic field has been detected \citep[see e.g.\,][]{Schleuning98, Girart06, Girart09, Qiu14, Kandori17, Maury18, Beltran19, Redaelli19, Kwon19}. This configuration implies that either magnetic lines efficiently resist the angular momentum due to rotation or that the latter is weak, otherwise the field lines would show a clear toroidal component. However, there are few cores where a spiral pattern suggests that the kinetic energy associated to angular momentum may dominate  \citep{Lee19, Beuther20,Sanhueza21}.  In some cases \citep{2017ApJ...842L...9H}, the observations reveal quite disorganized B-field topologies, which suggest weak-field conditions, and the possible role of turbulence or anisotropic gas flows in organizing the magnetic field lines. 

Moreover, there are two interesting features that ALMA observations have revealed. One is the detection of significant polarization along the outflow cavity walls, as in some cases this is the only place where polarization is indeed detected \citep{LeGouellec19, Hull20}. The other is the presence of well organized polarized filaments with the magnetic field along the filament \citep{LeGouellec19, Takahashi19, Hull20}. This may be  related with accretion streamers that have been detected recently \citep{Yen19, Alves20, 2020NatAs...4.1158P}, although kinematic information is needed to confirm this scenario.

\citep{Koch14} compiled the magnetic fields properties of 50 low and high mass protostellar cores. They propose four types of morphology \citep[see also][]{Koch13}: (1) magnetic field mostly aligned along the minor axis of the core, (2)  magnetic field mostly aligned along the major axis of the core, (3) hourglass and quasi-radial fields, and (4) irregular shapes.  This correlation was also confirmed by \citep{Zhang14}, which also found that the magnetic fields in massive dense core scales are either parallel or perpendicular to the parsec--scale magnetic fields.  However, there are cases where this correlation at different scales does not hold \citep{Girart13, Hull17}. Despite the overall trends where magnetic fields appear to be coherent with respect to larger scales and core orientation, there is no correlation with molecular outflow direction \citep{2013ApJ...768..159H, Zhang14}. However, there are other works that found some correlation between the outflow direction and the magnetic field, although the alignment is far from being perfect \citep{Galametz20, Yen21}. These two works show that, on one hand, the misalignment between the outflow and magnetic field direction appears to be correlated with the amount of angular momentum \citep{Galametz20}, and, on the other hand, the observed misalignment is not sufficient to reduce the efficiency of magnetic braking \citep{Yen21}.  

Most of the observations have relied on indirect methods to derive the magnetic field strength from the (mostly dust) polarization observations (see Section~\ref{obs-Bsignatures}). In spite of not being very accurate, these methods give a good approximation of the magnetic field strength. Thus, comparison between semi-analytical models or MHD simulations with polarization observations have found good agreement with the values obtained using different Davis-Chandrasekhar-Fermi approximations \citep{Goncalves08, Frau11, Juarez17, Maury18, Beltran19}. Overall, the DCF shows that star forming cores appear to be supercritical by a factor of about 2, and that these cores are gravitationally bound \citep{Myers21}.
Measuring directly the (line-of-sight) strength of the magnetic field is very difficult, and there is a small amount of positive detection at core scales \citep{Crutcher99a, Falgarone08, Crutcher09a, Pillai16, Nakamura19}. DR21(OH) is probably the only core where three different techniques have been used to derive the field strength \citep{Crutcher99a, Hezareh10, Girart13}. For the few sources with magnetic field strength derived from both (OH) Zeeman and dust polarization, the ratio of the plane-of-sky magnetic field component to its line-of-sight component, $4.7\pm2.8$, is larger than the expected average value for random orientations \citep{Myers21}. However, we should note that the OH Zeeman observations probably trace lower densities than (sub)mm dust observations. In any case, using both data sets, the magnetic fields appear to increase with volume density as $B \propto n^{2/3}$, indicating that magnetic field strength is significant but not enough to avoid the gravitational collapse \citep{Myers21}. 
Finally, it is important to be aware that the interpretation of the observational data is sometimes a source of vivid debate \citep{Mouschovias10, Crutcher10, Tassis14}, and the application at core scales should be taken with caution \citep{Reissl21, LiuJ21}.

\subsection{Protostellar disks}
\label{ssec:disks-obs}

One of the major predictions of magnetized models regards the properties of rotationally-supported disks: embedded young disks are expected to be compact and dust-rich. In this section, we briefly summarize recent constraints brought by observations of the sizes and masses of protostellar disks.

\begin{figure*}[ht]
\begin{center}
\includegraphics[width=0.7\linewidth]{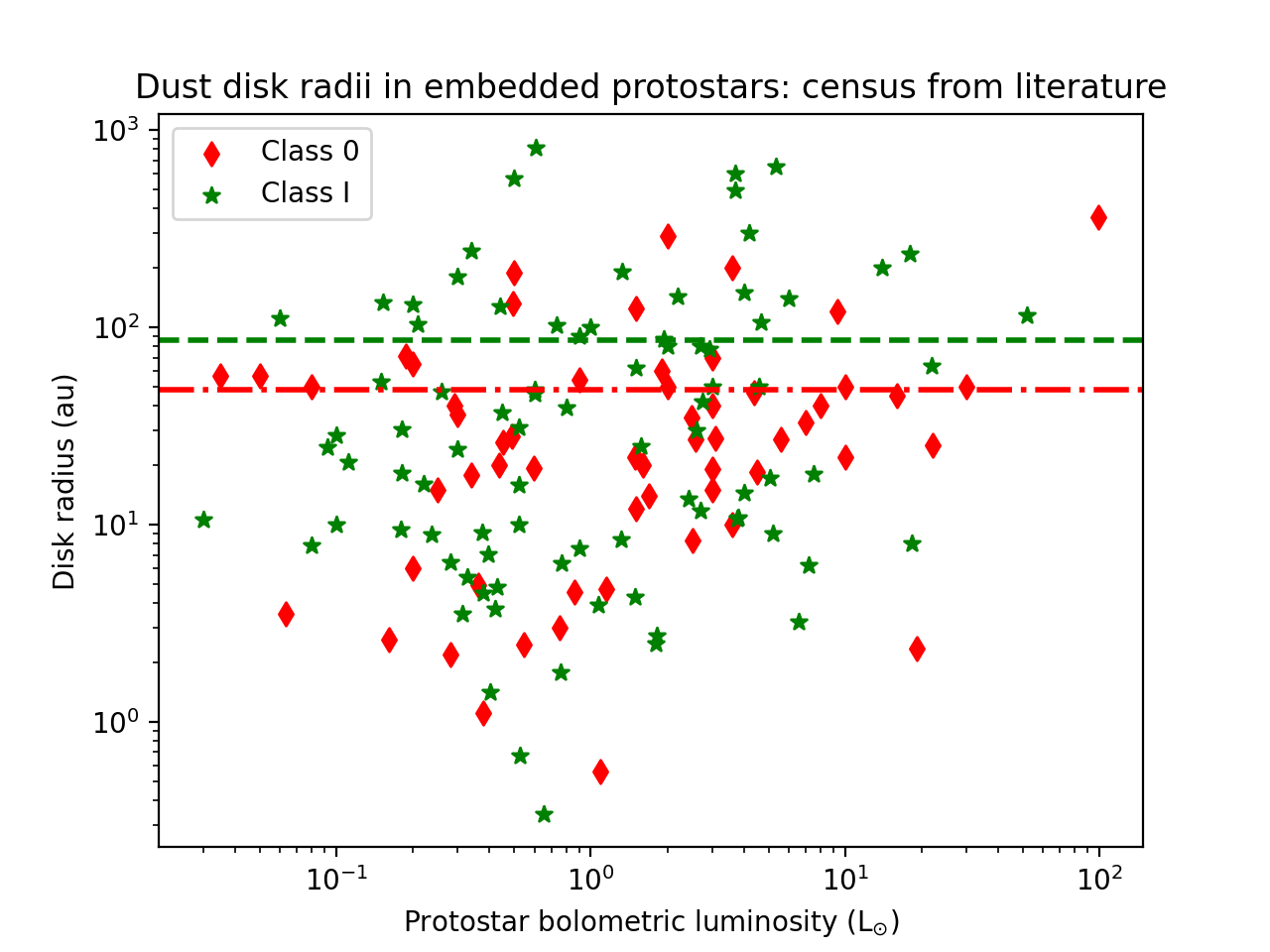}
\end{center}
\caption{Class 0 (in red) and Class I (in green) protostellar disk radii observed, from the dust continuum emission at millimeter wavelengths ($\lambda<2.7$mm). Most measurements stem from the CALYPSO survey \citep{2019A&A...621A..76M} and the VANDAM survey \citep{Sheehan22}. Other sources from \citet{2014ApJ...795..152O, 2018ApJ...863...94L, 2015ApJ...812..129Y, 2017ApJ...834..178Y, 2017ApJ...851...83C, 2015ApJ...812...27A, 2014ApJ...796..131O, 2013A&A...559A..82B, 2014A&A...562A..77H, 2014A&A...567A..32M, 2017ApJ...851...45S,2021ApJ...913..149E}. The mean dust disk radii for each Class are indicated as horizontal lines (48 au for Class 0 and 86 au for Class I).}.
\label{fig:Rdisks}
\end{figure*}

\citet{2010A&A...512A..40M, 2019A&A...621A..76M} used the NOEMA CALYPSO survey, at 1.3\,mm and 2.7\,mm to characterize the disk properties in twenty-six Class 0 and I protostars. Modeling the millimeter dust continuum emission with a combination of envelope and disk contributions directly in the visibility space, they find an average disk size of $<$ 50 au $\pm$ 10 au in the Class 0 objects, and 115 $\pm$ 15 au in the Class I objects. 
The VANDAM survey \citep{2018ApJ...866..161S,2020ApJ...890..130T} used VLA then ALMA observations of the mm dust continuum emission to characterize the radii of Class 0 and Class I disk sizes in Perseus and Orion. Their most recent results in Orion performing multi-wavelength analysis (25 Class 0 sources, 44 Class I sources) report smaller median dust disk radii than their preliminary analysis, with Class 0 disks having radii $\sim 35$ au in Orion \citep{Sheehan22}, and  no statistically significant difference between the properties of Class 0 and Class I disks. This later point may be specific to Orion, however, as our census of all embedded disks radii from the literature presented in Figure \ref{fig:Rdisks} does show a difference, with Class I disks being statistically more extended.
\citet{2017ApJ...851...83C} and \citet{2021ApJ...913..149E} used 2D Gaussian fits applied to  the ALMA dust continuum images of Class I protostars in Ophiuchus and find respectively a median dust disk radius from 12.6 au to $\sim 23.5$ au. These measurements are also included in Figure \ref{fig:Rdisks}.

These aforementioned studies rely mainly on the analysis of the thermal dust emission, but protostellar disks can also be identified thanks to kinematic signatures. Inside the centrifugal radius, the gas rotates with nearly-Keplerian motions as the pressure and centrifugal acceleration balance the radial gravitational acceleration. Measuring the size of rotationally-supported disks in embedded protostars needs to distinguish the disk emission from that of the infalling envelope, kinematically. 
Identifying the transition from envelope kinematics dominated by infalling gas ($v_r \propto r^{-1/2}$ and  $v_r > v_{\phi}$) to gas contained in a rotationally supported disk, with $v_{\phi}(\rm{r}) \propto \sqrt{(G \times M_{\star+disk})/r}$ and $v_r << v_{\phi}$, is thus, in theory, a simple way of measuring disk gaseous sizes. In practice, the simultaneous contribution of gas kinematics from different origins, chemistry effects on molecular gas tracers, and projection effects of asymmetric gas motions make it a hard task to clearly measure the rotation velocity $v_{\phi}$ in protostellar envelopes down to the disk outer radius. Only few protostars show clear signatures of Keplerian motions that can be used to characterize their disk: in the few objects that have an estimate of the dusty disk sizes, the gaseous disk size and dusty disk sizes are similar, within the relatively large uncertainties associated \citep{2014ApJ...796..131O, 2017ApJ...834..178Y,2019A&A...631A..64B,2020A&A...635A..15M, 2014A&A...562A..77H, 2014ApJ...796...70C, 2014ApJ...793....1Y, 2015ApJ...812...27A}.

Protostellar disk masses are usually measured from the dust emission, estimating dust masses from a set of hypothesis on the dust properties then correcting by a gas-to-dust ratio to obtain gas masses. The choice of wavelength(s) for the dust observations is crucial: for example, in Perseus, \citet{2020A&A...640A..19T} measure median dust masses of the embedded disks $\sim 5\times10^{-4} \msun$ for Class 0 and $2\times10^{-4} \msun$ for Class I from the VLA dust continuum emission at centimetre wavelengths. In comparison, dust masses from sub-millimetre dust emission in ALMA bands towards the same population of disks in Perseus are significantly lower, $1.4\times10^{-4} \msun$ and $4 \times10^{-5} \msun$ for 38 Class 0 and 39 Class I respectively. These lower dust disk masses are in better agreement with the dust masses found in Orion embedded disks: it suggests that dust masses extrapolated from long wavelengths may be significantly over-estimated, due to the unknown emissivity and size distribution of dust grains in the dense environments that are disks. In Ophiuchus, the mean Class I disk dust mass is found to be significantly lower, $9 \times10^{-6} \msun$ \citep{2019ApJ...875L...9W, 2021ApJ...913..149E}: it is still about 5 times greater than the mean Class II disk mass in the same region, but the dispersion in each class is so high that there is a large overlap between the two distributions. \cite{Sheehan22} used multi-wavelength analysis of the SED towards Orion protostars. These estimates are also associated to large error bars, but the multi-wavelength analysis they performed should allow to better discriminate the disk from the envelope contribution and hence provide more robust results. They find median dust mass $2 \times10^{-5} \msun$ for the Class 0 protostars, and $1.5 \times10^{-5} \msun$ for the Class I protostars, pointing towards much smaller embedded disk masses and less clear mass differences between the evolutionary stages than suggested by previous simplistic estimates made from single wavelength analysis.
Finally, when the temperature of embedded disks can be investigated, observational studies suggest these disks may be warmer than their older counterparts, with typical temperatures around 20-30 K \citep{2020ApJ...901..166V,Zamponi21}.

For the high mass stars, the rapid dynamical evolution and the larger distances make the detection of the rotationally supported disk more difficult, since they are in most cases surrounded by very massive molecular envelopes. This makes the identification of the disks more controversial. Resolving their velocity structures, masses, and sizes has been made possible only recently thanks to ALMA. Relatively isolated disks are very rare, such as Orion~I, which has unique features possibly related to a violent multiple star interaction event \citep{Plambeck16, Hirota17, Ginsburg19, Wright22}. 
Other clear cases of disks around massive stars, that appear to be a scaled up version of the ones around low-mass stars, are GGD~MM1 \citep{Girart18, Anez20}, G11.92-0.61~MM1 \citep{Ilee18}, and G17.64+0.16 \citep{Maud19}, with very massive (2--5~$\msun$), dense and hot ($\gtrsim $400~K) disks. 
However, there are many cases reported in the literature where Keplerian velocity patterns can be fitted towards disk-like structures, which can extend from few hundreds to a thousand au \citep[e.g.][]{Sanchez13, Johnston15, Cesaroni17, Beuther17, Girart17, Tanaka20, Williams22}. In most cases, a detailed measurement of the gas temperature and surface density is needed to  check the stability of such structures. The presence of disks around massive stars has also been found through near/mid-IR interferometry  \citep{Kraus10, Frost19}. 

These two works suggest that the observed massive disks are a scaled up version of low-mass disks, but need to be confirmed with larger samples. 

\subsection{Fragmentation into multiple systems}
\label{sec:protosystems-obs}

Measuring the multiplicity of  Class 0 and I solar-type protostellar systems has been challenging and the focus of many observational works. 
Submillimetre/millimetre observations are the only reliable tool for characterizing Class 0 and I multiplicity, as near-infrared emission towards embedded protostars suffer from significant extinction due to the envelopes, and contamination by scattered light, while near-infrared observations are usually robust at characterizing multiplicity in more evolved YSOs \citep{Duchene13}.
The observed faction of binary and multiple systems in low mass protostars are high, and similar to the multiplicity of main sequence stars in the field. 
On average, a multiplicity percentage of 64\% is found among Class 0 protostars with linear separations in the range 50 - 5000~au (e.g., \citealt{Looney00, 2010A&A...512A..40M, Enoch11, Tobin13, Bouvier21}), whilst it is between 18\% and 47\% for Class I sources with linear separations in the range 45 - 5000 au \citep{Haisch04, Duchene04, Duchene07, Connelley08a, Connelley08b}. For example, in the Orion molecular cloud, observations of 300 Class 0, I, or Flat Spectrum YSOs found a total of 85 multiple systems  at separations $\simlt 10000$~au, 58 multiples at separations $\simlt$1000~au, and only 47 multiples with maximum separations less than 500 au \citep{Tobin22}. No statistically significant difference in the separations with evolutionary stages during the protostellar stages were found in the most complete studies so far, while tentative evidence is found that field stars and non-embedded YSOs may feature a lower number of wide (e.g., $a>1000$ au) multiple systems than their embedded counterparts, in Orion \citep{Tobin22}. Note that this result may be significantly driven by incompleteness at sampling wide systems around the most evolved YSOs and field stars: this limitation has now been partially lifted thanks to increasing astrometric precision and led, for example, to the discovery of a new large population of wide systems in Taurus within the range 1–60 kAU \citep{Joncour17}. Finally, no statistically significant difference was found regarding the multiplicity fractions observed between regions of high and low YSO density, also in Orion.

\begin{figure*}[ht]
\begin{center}
\includegraphics[width=0.7\linewidth]{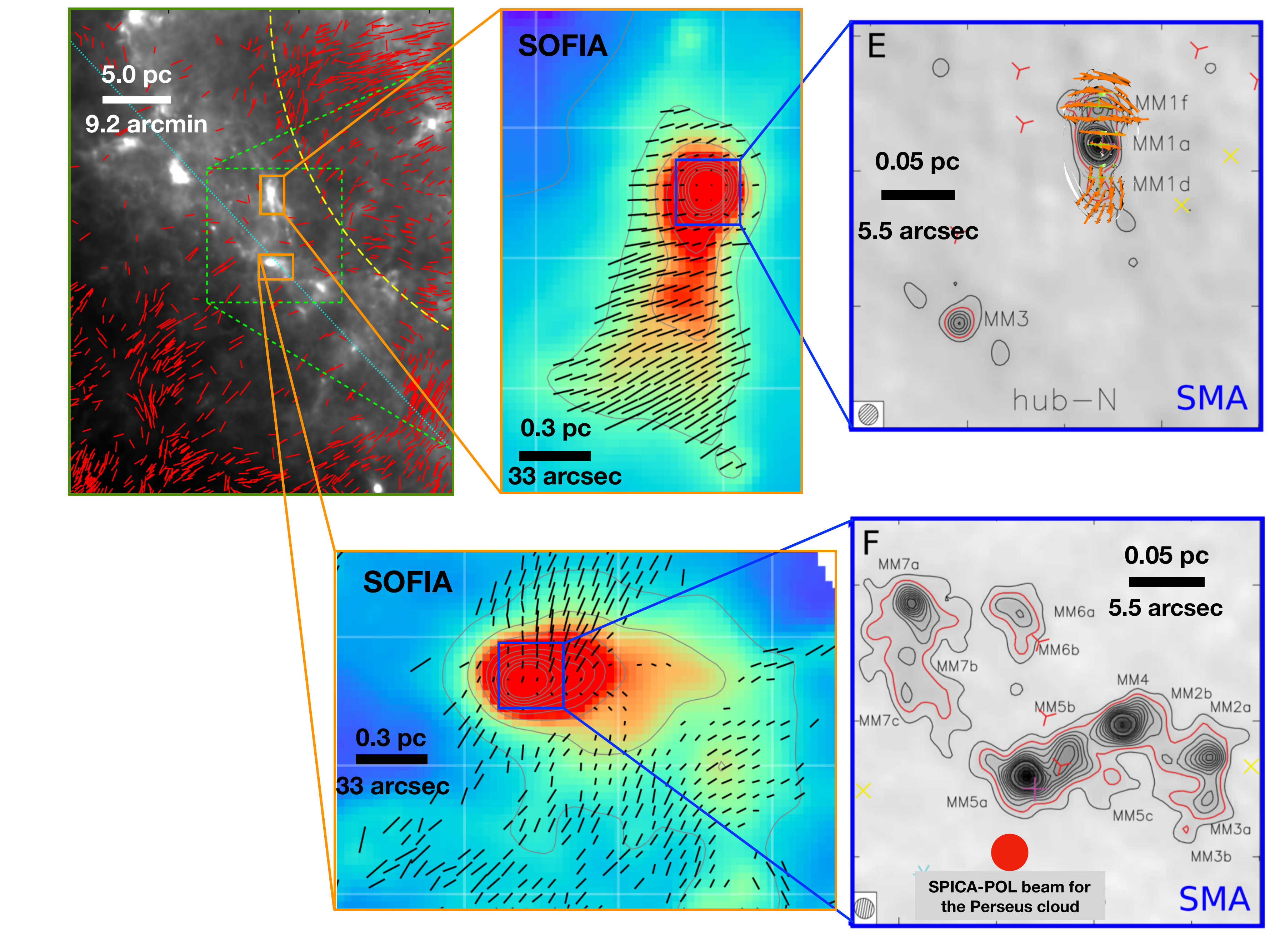}
\end{center}
\caption{Composite images of the G14.225-0.506 massive star
forming region \citep{Busquet16,Santos16}.
Top left panel: R band optical polarization vectors (red segments)
overlaid on Herschel $250 \mu$m image overlapped (from \citealt{Santos16}). Central panels: SOFIA/HAWC$+$ $200 \mu$m images (beam
$14\arcsec$) of the Northern (top) and Southern (bottom) hubs, with
black segments showing the magnetic field direction (F. Santos,
private communication). Right panels: Submillimeter Array (SMA)
images of the 1.2 mm emission toward the centre of the Northern
(top) and Southern (bottom) hubs \citep{Busquet16}, with
orange segments showing the magnetic field direction (A\~nez et al. in prep.).} 
\label{fig:G14}
\end{figure*}

Massive stars have a predilection for forming in clustered environments with other protostars, and their young embedded counterparts are thus often studied in cluster star-formation modes \citep{Cyganowski17}.
High mass star forming cores observed at high angular resolution appear to show different level of fragmentation \citep{Beuther04, Rodon12, Palau13, Busquet16, Sanchez17, Liu19, Sadaghiani20}. The process of fragmentation is hierarchical \citep{Beuther04, Beuther15} and appears to proceed down to disk scales \citep{Beuther17, Busquet19}.
The first systematic study of the level of fragmentation down to scales of $\sim 1000$~au suggested that this could be set by the initial magnetic field/turbulence balance \citep{Palau13}. Subsequent studies found that the fragmentation level was only mildly correlated with the density profile, and that it is consistent with thermal or turbulent Jeans fragmentation \citep{Palau14A, Palau15, Palau18, Beuther18, Liu19}.
However, in a more recent work when the fragmentation level is compared with the magnetic field properties derived from the dust polarization, there is a tentative correlation of the fragmentation level with the mass-to-flux-ratio \citep[see Fig.~5, and][]{Busquet16, Nacho20, Palau21}.

\section{Discussion: constraints on magnetized models from the observations}
\label{sec:discussion}

\subsection{Magnetic fields}

The observations of (sub-)millimeter polarized dust emission show that the magnetic field is detected in all dense environments producing stellar embryos. Assessing whether the observations can be trusted to infer statistically robust constraints on the magnetic fields threading protostellar cores, and compared to B-fields in models, requires the analysis of synthetic observations from magnetized models.
\citet{LeGouellec20} and \citet{Valdivia22} have post-processed outputs from non-ideal MHD models of protostellar evolution with the \textsc{Ramses} code \citep{Fromang06} (see also the work of \citealt{kuffmeier2020} in Figure \ref{fig:synthobs-kuffi}). 
Their work analyses the resulting synthetic polarized dust emission maps, 
which are compared to the true B-field properties in the models and to observations.
\citet{Valdivia22} show that measurements of the line-of-sight averaged magnetic field line orientation using the polarized dust emission are precise enough to recover the mean field lines distribution with accuracy $<15\deg$ (typical of the error on polarization angles obtained with observations from large mm polarimetric facilities such as ALMA) in about $95 \%$ of the independent lines of sight peering through protostellar envelopes at all radii. 
When focusing on the smaller envelope radii $<500~\mathrm{au}$, where the magnetic field lines are more likely perturbed from the initially smooth configuration by infall and outflow and opacity effects kick in, $75 \%$ of the lines of sight still give robust results.
Finally, \citet{Valdivia22} and \citet{LeGouellec20} find that the polarization fraction is correlated to the degree of organization of the magnetic field along the line of sight, confirming that the behavior observed at larger scales, in less dense gas, holds true down to the scales of the envelopes of protostars. 
\citet{LeGouellec20} find that the dust alignment efficiency does not significantly vary with local gas density in their observations, and that the synthetic maps only can reproduce the observed polarimetric data when significant irradiation from the central protostar is included. \citet{Lam21} have used numerical models compared to ALMA observations of the polarized dust emission by \citet{2018ApJ...855...92C}: comparing polarization fractions they also conclude that dust polarization at ALMA wavelengths is most likely due to magnetically aligned grains in inner envelopes and dust scattering in disks.

\begin{figure*}[!h]
\begin{center}
\includegraphics[width=0.8\linewidth]{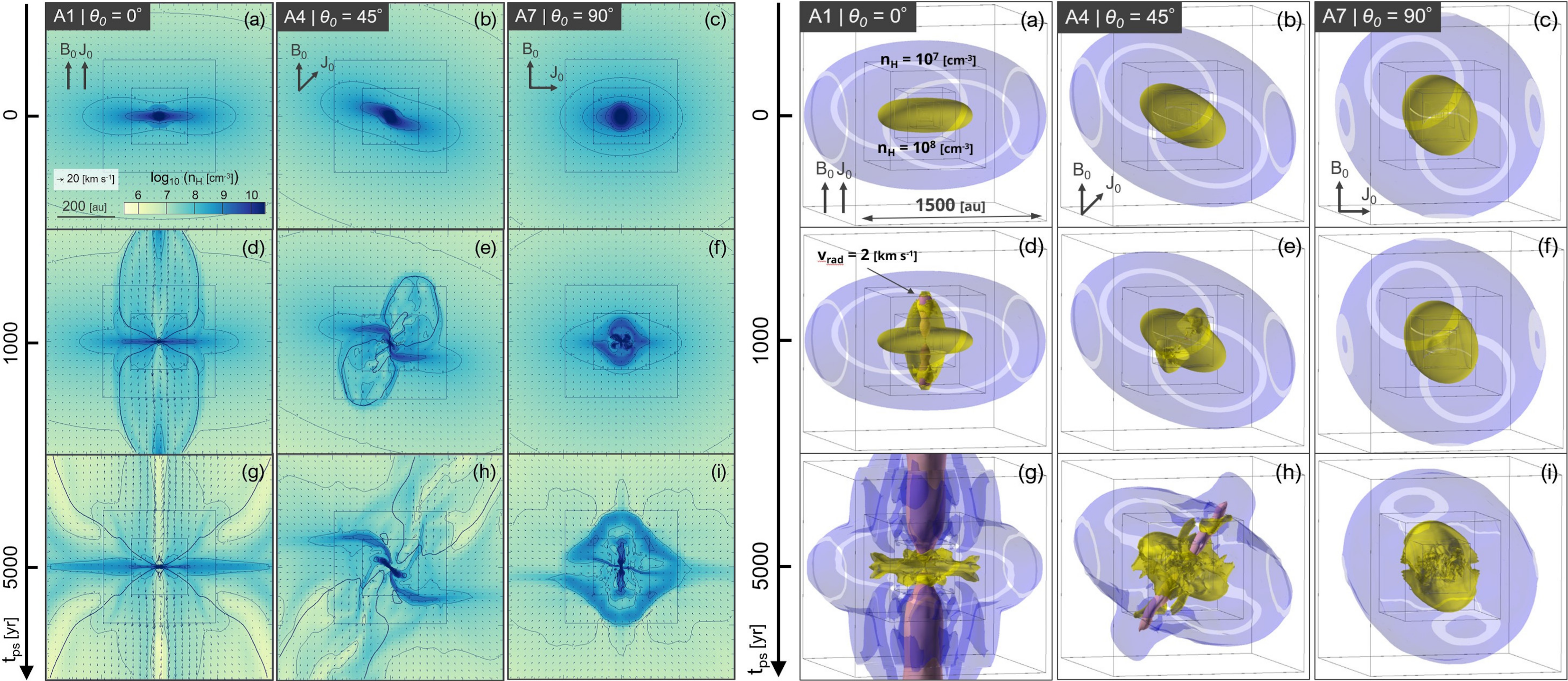}
\end{center}
\caption{Left: Density (color and contours) and velocity (arrows) distributions in models from \citet{hirano2020}, implementing different angles between the magnetic field initial direction and the initial rotation axis of the core, at $t_{\rm ps} = 0$, $1000$, and $5000$\,yr after protostar formation. The box size is $780$\,au.
Right: Three-dimensional structures for the same models, highlighting two isodensity contours in purple and yellow, and one isovelocity contour of radial velocity $v_{\rm rad} = 2\,\kms$ (in red). Note the absence of an outflow in the case of orthogonal configuration (third column): this case thus seems unrealistic considering the prevalence of outlows observed around low-mass protostars.}
\label{fig:B_J_models-hirano}
\end{figure*}

Observationally, the correlation between outflow axis and the mean magnetic field direction has received a lot of attention because it is believed that the efficiency of magnetic braking to redistribute angular momentum and prevent the growth of disks to large radii depends on the configuration of the core's magnetic field with respect to the core's rotation axis when the collapse starts \citep{2012A&A...543A.128J,hirano2020}, as portrayed in Fig. \ref{fig:B_J_models-hirano}. 
Using an SMA survey of twenty low-mass protostars, \citet{Galametz18, Galametz20} has found that the protostellar envelopes tend to have a higher angular momentum associated to rotation on 5000 au scales if the mean envelope magnetic field measured at similar scales is misaligned with the rotational axis of the core, assumed to coincide with the outflow axis. On the other hand, observations analyzed in \citet{2021ApJ...916...97Y} have shown no correlation of the dust continuum protostellar disk radii with the misalignment between the magnetic fields and outflow axes in Orion A cores. Protostellar gas kinematics from the SMA MASSES (sample of 32 protostellar envelopes in Perseus) survey brings further support to the scenario proposed by \citet{Galametz20}, with a significant correlation between the rotational velocity gradients (at 1000 au, normalized by the infalling velocity gradients) and the misalignment angles between the magnetic fields and outflows \citep{Gupta22}. At the very small envelope radii, the mean B-field direction are observed to be randomly aligned with respect to the outflow axis \citep{2013ApJ...768..159H}. However, as reported by \citet{2014ApJS..213...13H}, there is evidence that cores with lower fractional polarization tend to have their outflows perpendicular to the mean B-field, which may suggest the existence of a non-negligible toroidal field morphology (caused by the core/disk rotation) in addition to the poloidal one caused by the envelope-disk accretion.

These results may be confronted to models where the magnetic field is efficient at reducing the amount of angular momentum transmitted to the inner envelope scales ($< 1000$ au), inhibiting the formation of large hydro-like disks. In some cases, numerical simulations have shown tentative evidence of better alignment of core-scale B fields and outflows in the more magnetized models \citep{2017ApJ...834..201L}. However, these observational results may also support the hypothesis that the angular momentum responsible for the formation and growth in size of protostellar disks has a more local origin, at a few hundreds of au scales, as discussed in recent theoretical studies following the disk evolution and angular momentum evolution in star-forming cores \citep{2020A&A...635A.130V, 2021MNRAS.502.4911X, 2021A&A...648A.101L}. Future observations will undoubtedly bring more constraints on the dynamical role of B-fields to regulate the protostellar collapse and set the pristine disk properties, with the upcoming large surveys thanks to the development of polarimetric capabilities and tools to model, e.g., polarized dust emission. Observations, with e.g. the large radio interferometers and next generation of mid-infrared facilities, of the small scale structures of these young disks, still largely unresolved spatially, will also shed light on this question.

\subsection{Coupling the magnetic field to protostellar material}

\begin{figure*}[!h]
\centering
\includegraphics[width=0.6\linewidth]{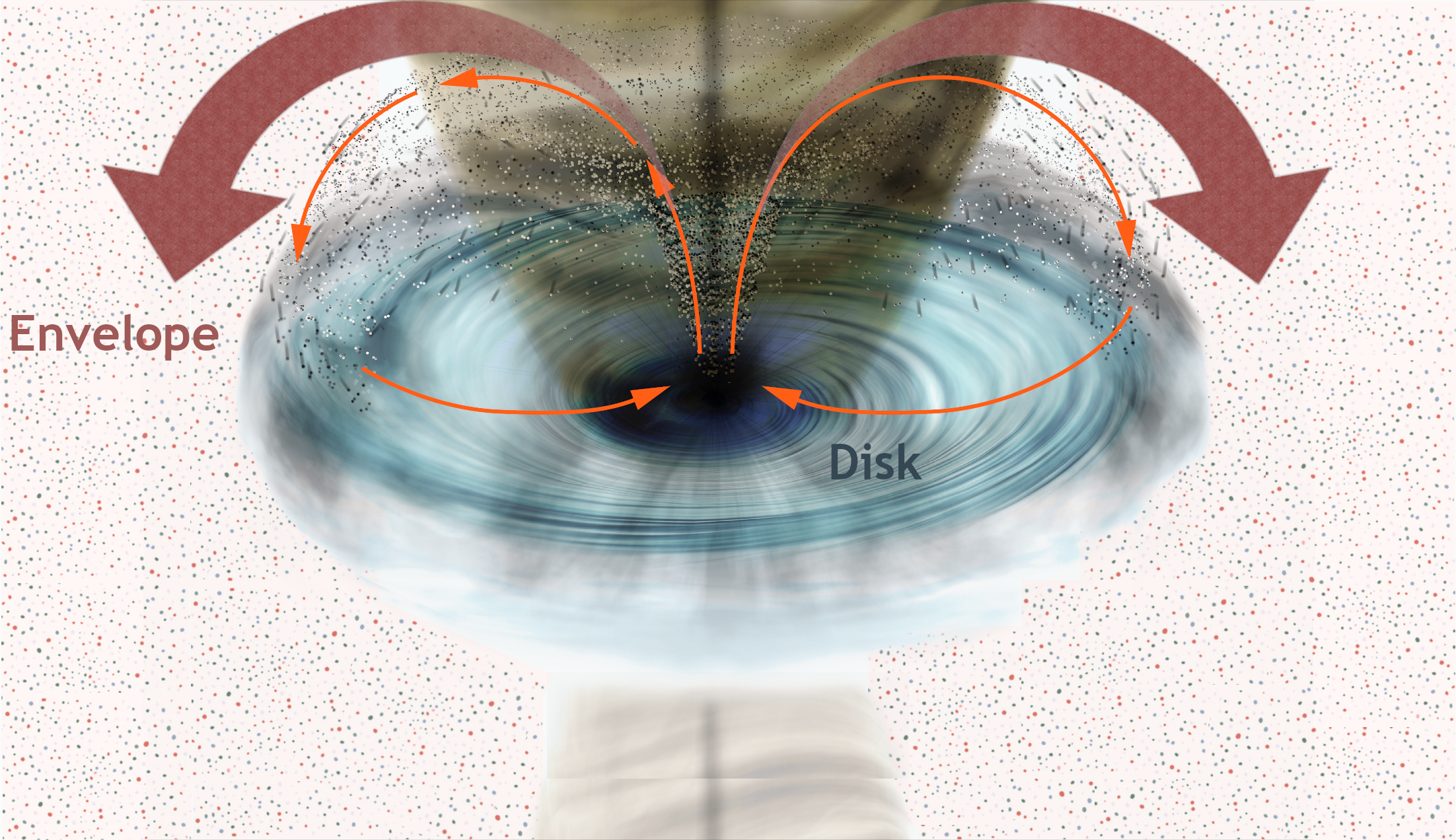}
\caption{Artist impression of the dust growth and recycling from disk to envelope thanks to outflowing gas during the protostellar stage. Modified from an original illustration in \citet{Tsukamoto21}}
\label{fig:tsukamoto_dust}
\end{figure*}

Observations of the low dust emissivities and high polarization fractions at mm wavelengths in embedded protostars suggest that partial grain growth, maybe up to grain sizes $>100$ \mic\, could have already occurred during the first $0.1$ Myr of the star formation process.
If these low emissivities and high polarization fractions  do indeed trace a population of large dust grains in the inner layers of protostellar envelopes less than 0.1 Myrs old, the timescales to grow grains up to sizes $>100$ \mic\ in dense star-forming material may need to be revised. For example, considering porous dust grains may help in growing grains to sub-millimeter sizes in a few dynamical timescales during the protostellar collapse \citep{2009A&A...502..845O}. Also, since small grains are not only coupled to the gas but also to the magnetic field, a magnetized collapse may also help segregating the grains with different charges, and ultimately shortening the grain coagulation timescale \citep{Hoang21}. 
Also, the presence of rather large grains at early stages in the protostellar envelopes may increase the ability of  protostellar disks to be efficient forges to build up even bigger dust grains, that are re-injected in envelopes thanks to protostellar winds and outflows \citep{Wong16,2017MNRAS.465.1089B, 2020A&A...641A.112L,2021ApJ...907...80O,Tsukamoto21}. Figure \ref{fig:tsukamoto_dust} shows an artist view of such circulation processes potentially affecting dust grains during the protostellar phase.

Moreover, observational evidences of grain growth during the embedded stages also bear strong consequences on the efficiency of magnetic fields to regulate the transport of angular momentum.
Indeed, whereas in the diffuse interstellar medium, electric charges are carried by electrons and protons, in the dense cores and particularly at high densities, the small dust grains are the main charge carriers \citep{nishi1991,nakano2002,zhao2016}.
The disappearance of small dust grains while forming bigger grains increases magnetic resistivities 
\citep{nishi1991,Guillet20} and consequently let more B flux leak outwards during the collapse of the gas from the envelope onto the star-disk system \citep{2021MNRAS.505.5142Z,2020A&A...643A..17G}. Hence the disk properties may also depend strongly on the dust properties in the inner envelope.

\begin{figure*}[ht]
\begin{center}
\includegraphics[width=0.6\linewidth]{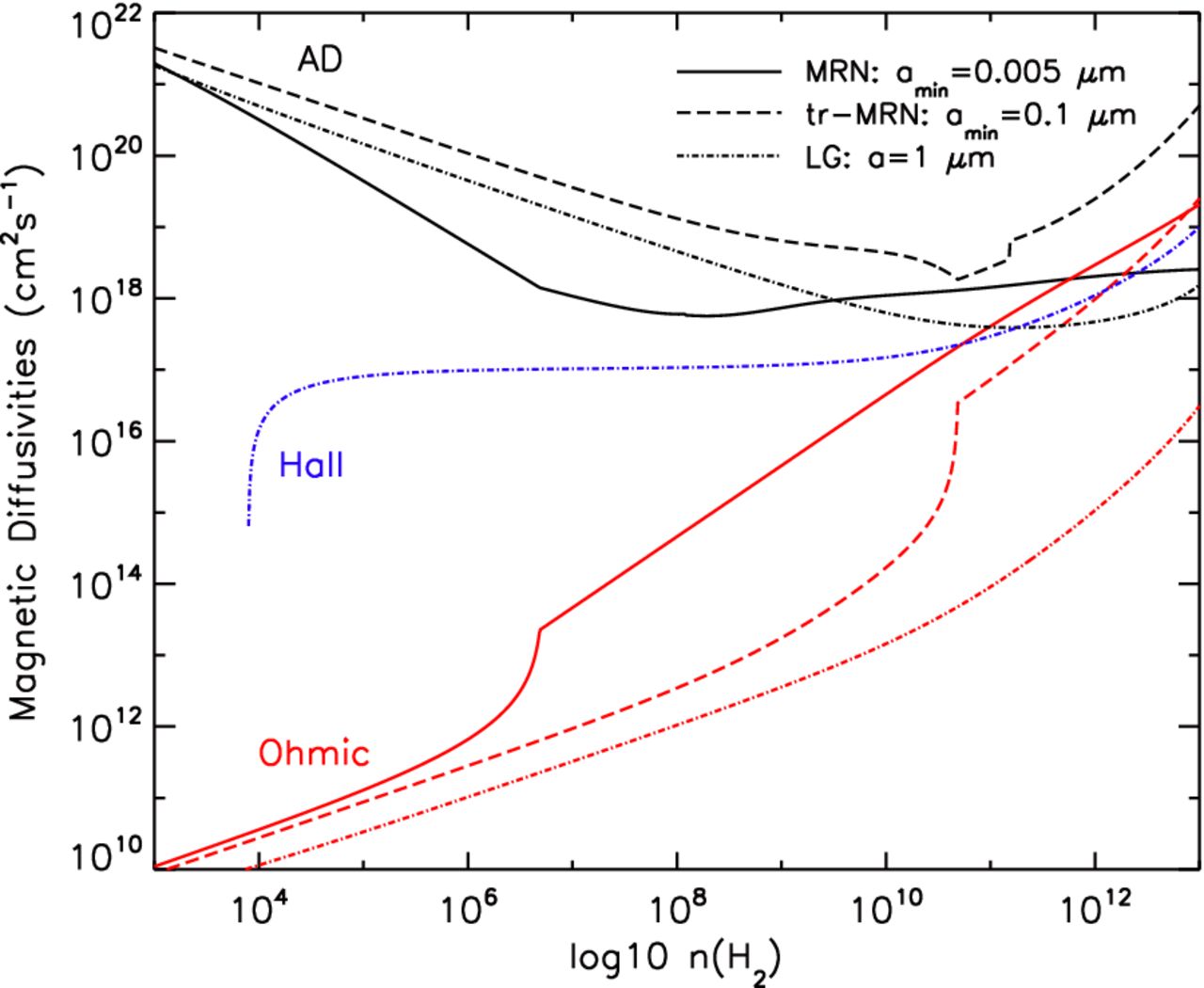}
\end{center}
\caption{Change of MHD resistivities when removing the very small grains, from \citet{zhao2016}.}
\label{resist-dust}
\end{figure*}

The influence of the grain distribution on disk 
formation has been investigated through numerical 
simulations by \citet{zhao2016}. By removing the population 
of very small grains, the authors conclude that the ambipolar 
diffusion is enhanced by 1-2 orders of magnitude, see Figure \ref{resist-dust}. As expected the numerical simulations reveal that indeed the centrifugally 
supported disks, which form, sensitively depend on the 
presence of the very small grains. For instance, for a particular set of parameters (with a relatively strong field aligned with the rotation axis), no disk would form when a MRN grain distribution is assumed while disks of several tens of AU radius form when a truncated MRN distribution is employed. Let us stress that the presence of very small grains in dense cores is presently poorly constraint. From a theoretical point of view, \citet{2020A&A...643A..17G} have shown that the latter may be efficiently removed by coagulation 
due to the drift between different dust species induced by 
ambipolar diffusion \citep[see also][]{silsfbee2020}. To what extent this population could not be replaced, for instance by fragmentation of bigger grains, remains to be investigated. At disk scales, where the large densities are favorable for rapid dust growth, up to $a_d\sim 1 \mm$, the magnetic resistivity can become many orders of magnitude bigger if small grains are rare, weakening ambipolar diffusion and recoupling the magnetic field and the gas.

Resistivities are also strongly dependent on the ionisation rate. Whereas most of the calculations assumed a value 
of a few $\zeta=10^{-17}$ s$^{-1}$ inferred from typical 
cosmic ray galactic abundances \citep{padovani2009}, more extreme ionisation rates and their consequences on disk formation,
have been recently explored.  \citet{wurster_cosm_2018} 
performed simulation with $\zeta$ varying from $10^{-13}$
to $10^{-24}$ s$^{-1}$. They concluded that for 
$\zeta > 10^{-14}$ s$^{-1}$, the results are essentially identical to ideal MHD while for $\zeta < 10^{-24}$ s$^{-1}$, 
they are close to hydrodynamics.
\citet{kuffmeier2020} specifically explored the disk 
radius and mass dependence on the ionisation rate, varying it by a factor of 10 from $\zeta=10^{-17}$  to 
$\zeta=10^{-16}$ s$^{-1}$. Performing bi-dimensional simulations
with a mass-to-flux of 2.5, they found that with the 
highest $\zeta$, no disk forms while a few tens of AU one
forms with their smallest $\zeta$. They further propose
that cosmic ray abundances may control disk mass and size and 
explain the differences between disks observed 
in various regions \citep[e.g.][]{cazzoletti2019}.

\subsection{Gas kinematics: rotation, mass infall and outflow rates}

\begin{figure*}[ht]
\begin{center}
\includegraphics[width=0.8\linewidth]{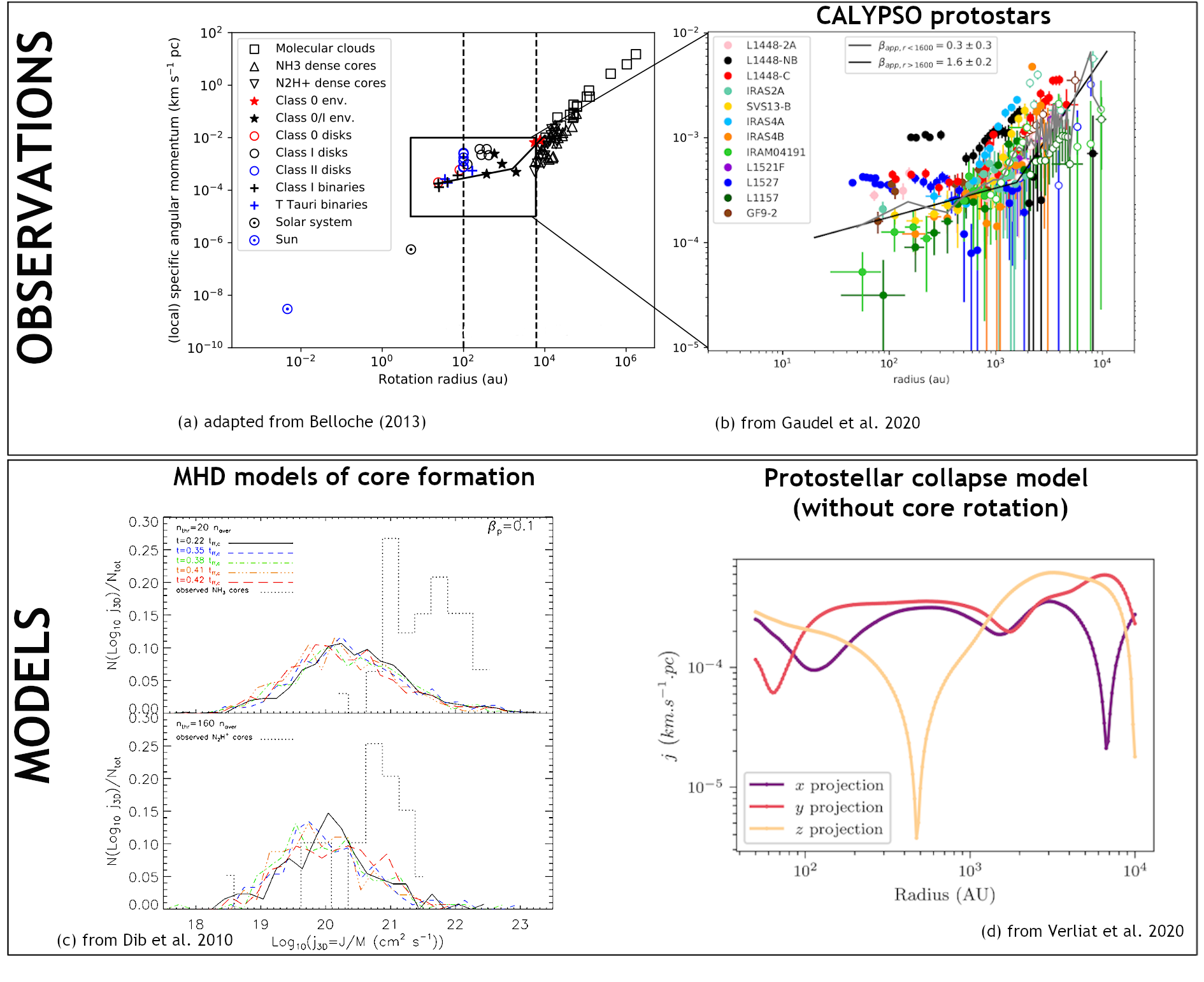}
\end{center}
\caption{
Observations and models of the angular momentum at core scales and disk scales, from \citet{Belloche13a,2020A&A...637A..92G,Dib10,2020A&A...635A.130V}.}
\label{fig:angmom}
\end{figure*}

Observations do not suggest that cloud rotation is transmitted to star-forming cores \citep{Tatematsu16,Hsieh21}. 
At core scales, both hydrodynamical and magnetized models of core formation and evolution have shown that they do not produce rotating cores with the high angular momentum values measured by e.g. \citet{Goodman98} at 0.1pc scales. Recent observations and models suggest that the angular momentum measured at core scales (e.g. $\sim 5000$ au) could be due to non-axisymmetric motions associated to, e.g., turbulent and gravitational processes \citep{Burkert00,2008ApJ...686.1174O,Dib10,Kuznetsova19,2020A&A...635A.130V}.
Hence, large scales velocity gradients in envelopes should not be interpreted automatically as organized rotation: for characterizing the motions responsible for the angular momentum observed, and as illustrated in Figure \ref{fig:angmom}, resolved observations and their comparisons to model predictions are key. Finally, the angular momentum of the gas measured at envelope radii $<1000$~au would predict disk whose sizes are broadly consistent with the observed ones, although observations reveal a tentative decrease of $j_{\rm{spe}}$ in inner envelopes, which may trace a decrease of angular momentum at small radii \citep{2020A&A...637A..92G}. 
Observations of complex gas kinematics in protostellar envelopes, from disturbed core-scale velocity gradients to localized streamers (see Figure \ref{fig:streamers}) connecting the envelope to the disk scales confirm the wealth of observational evidence that the monolithic vision of collapse which prevailed in the past decades needs to be revised.

The rather high values of mass infall rates found in Class 0 protostars seem at first sight inconsistent with observed protostellar luminosities. While the bolometric luminosities of Class 0 protostars are typically a few $\rm{L}_\odot$\ (with large variations from object to object), such accretion rates should produce a accretion luminosities a few tens of $\rm{L}_\odot$\. This is the well known luminosity problem \citep{2012ApJ...747...52D}, which was proposed to be solved with a scenario of quiet accretion including sporadic large accretion bursts. It remains unclear how such rare events could be captured so efficiently by observations measuring the gas accretion rates, as a recent ALMA survey of protostars in Perseus suggest the frequency of the bursts decrease with protostellar evolution, from a burst every 2400 yr in the Class 0 stage to 8000 yrs in the Class I stage \citep{2019ApJ...884..149H}.
However, this scenario finds support in observations of molecular species: sudden temperature changes due to short but vigorous accretion bursts would explain why CO is observed where the current envelope temperature predicts it should be frozen onto dust grains ($<30$K) for example \citep{2016A&A...591A...3A,2016Natur.535..258C, 2017A&A...602A.120F}. It may also be that our lack of understanding of the physical conditions at the very small scales where most of the radiation is reprocessed to produce the observed luminosity, prevents us from properly model the expected luminosities, or that the gas mass infall rates measured at few hundreds of au in the envelopes are not representative of the true accretion rates at much smaller radii, onto the protostellar objects. Supporting this last hypothesis, it is interesting to note that MHD models following the evolution of low-mass cores find typical mass accretion rates $\sim 10^{-6} $ $\rm{M}_\odot$\ per year, with large variations \citep{hennebelle2020,Kuznetsova20}. The global mass accretion rate from the disk to the central star is lower when the disk structure is properly modeled, and the short-term variation (bursts) of the mass accretion is also highly attenuated with increased resolution \citep{2018MNRAS.475.2642K,2021A&A...648A.101L}. It seems therefore that either the protostellar accretion rates obtained from observations are over-estimated, which could be due to sampling the inner envelope and not the disk-star connection, or the models predict slower mass accretion than protostars really do accrete. We stress that the existence of preferential directions along field lines, and the development of highly non-axisymmetric collapse may be responsible, in magnetized models, to regulate the mass accretion rate directly from the inner envelope to the central star, without the need to transit through the disk. In their MHD models of the evolution of massive cores into protostars \citet{2021A&A...652A..69M} finds average mass accretion rates on the central object around $\sim 10^{-3}$ $\rm{M}_\odot$\ per year, which is also about an order of magnitude lower than estimates from the observations towards similarly massive protostars.

\subsection{The formation of disks}

\begin{figure*}[ht]
\begin{center}
\includegraphics[width=0.4\linewidth]{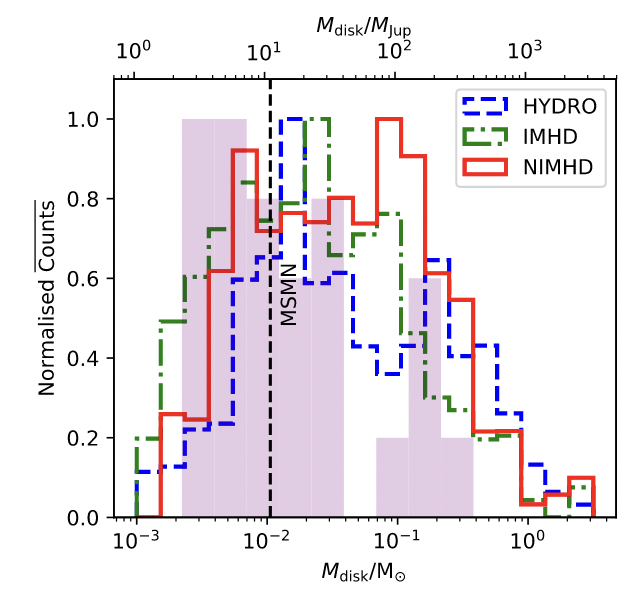}
\vspace{1cm}
\includegraphics[width=0.4\linewidth]{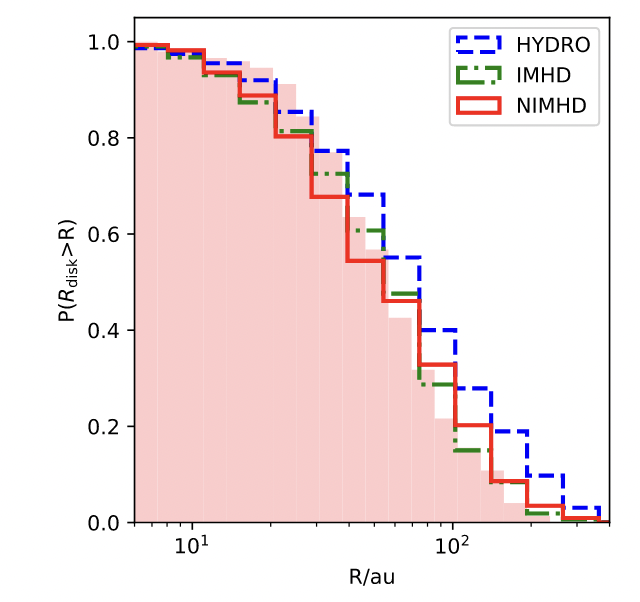}
\end{center}
\caption{Left panel: Distributions of disk sizes obtained for the three disk population synthesis models presented in \citet{lebreuilly2021} (note that the sizes are normalized so that their peak in the y-axis is 1). The mass of the miminum mass for the Solar nebula (MSMN) is indicated as a dashed line. Right: Cumulative distribution function of disk radius for the three models presented in \citet{lebreuilly2021}.}
\label{disk-sizes_lebreuil}
\end{figure*}

Recent observations benefiting from increased spatial resolution and sensitivities have firmly established  that most disks around embedded protostars are compact and not as massive as early studies had suggested. The current disk radii estimates from the literature are shown in Figure \ref{fig:Rdisks}. Such compact disks are difficult to produce in large fractions with purely hydrodynamical analytical models of disk formation conserving angular momentum, but could be a natural outcome of magnetized models of disk formation and models of anisotropic collapse \citep{hennebelle2020,Kuznetsova22}. 
Obtaining self-consistent disk populations from numerical simulations is quite challenging because it requires to be able to treat simultaneously spatial scales sufficiently small to resolve the protostellar disks while in the same time advance sufficiently the clump scales to allow the formation of a 
statistically significant number of disks. For this reason, only two studies have been reporting disk population self-consistently formed from numerical simulations. \citet{bate2018} 
presents Smooth particle hydrodynamical simulations
of a 500 $M_\odot$ clump
including radiative transfer, producing a disk population whose sizes 
ranges from about 10 to few 100 AU. The author compares 
the mass distribution inferred from simulations with several 
observational surveys and concludes that the disk formed in the 
simulations are up to 10 times more massive. 
\citet{lebreuilly2021} present three adaptive mesh refinement simulations of 1000 $M_\odot$ clump which treat 
 radiative stellar feedback and both ideal and 
non-ideal MHD. The spatial resolution is up to 1 AU.
Figure~\ref{disk-sizes_lebreuil} shows for the three runs, the mass distribution and the radius distribution, respectively.
For comparison, the data of VLA Class 0 disk mass distribution  \citep[purple,][]{2020A&A...640A..19T} have been reported, as well 
as the disk radius inferred by the Calypso
\citep{2019A&A...621A..76M} and VANDAM projects 
\citep{2018ApJ...866..161S,2020ApJ...890..130T}.
Altogether and as already suggested from individual studies that compared HD, MHD, and non-ideal MHD numerical models for single objects, the hydrodynamical disks appear to be bigger than the MHD ones and overall in less good agreement with observations than when magnetic field is included, particularly regarding the
disk radius distribution. Whereas there is some scatter in the distributions, these first population studies, properly taking into account the initial conditions, appear to be promising tools for future comparisons to observations.

A local origin of the angular momentum building rotationally-supported disks has been proposed by some models: for example, \citet{2020A&A...635A.130V} have shown that the observations of angular momentum in protostellar envelopes can be satisfactorily reproduced if the disk formation results from anisotropies in the local velocity field, and not a consequence of the transfer of angular momentum from organized large scale core rotation. As shown in Figure \ref{fig:angmom}, some observational evidence may support that scenario, with a good agreement between those models and the observations of the specific angular momentum found in envelopes, or more recently for example the observations of the L1489 protostar suggesting only the inner part of the envelope, at  radii $<4000-6000$, may be directly involved in forming the central star \citep{Sai22}. This would imply that the material participating directly to building the disk and the star contains much less angular momentum than the values measured at core scales, lessening the angular momentum problem for star formation.

Although it was not shown observationally in the most embedded disks, models suggest it may be that the big dust drifts even during the early disk phases, leading to different apparent disk sizes recovered at different wavelengths. The uncertainties regarding dust properties in these specific conditions also affect the dust masses and hence the disk masses that are estimated in these objects.
If observations suggesting young embedded disks are warm (20-30 K, \citealt{2020ApJ...901..166V,Zamponi21}) are confirmed, it could suggest gas kinematics within the disc play a significant role in heating the disk, as the protostellar radiation is expected to be highly extinct at such very high densities. For example, \citet{Zamponi21} compare observations with synthetic observations of MHD protostellar disk models formed after the collapse of a dense core, and suggest that heating due to gravitational instabilities in the disk is able to generate dust temperatures in agreement with observational constraints in the IRAS 16293 Class 0 disk.

\begin{figure*}[ht]
\begin{center}
\includegraphics[width=0.9\linewidth]{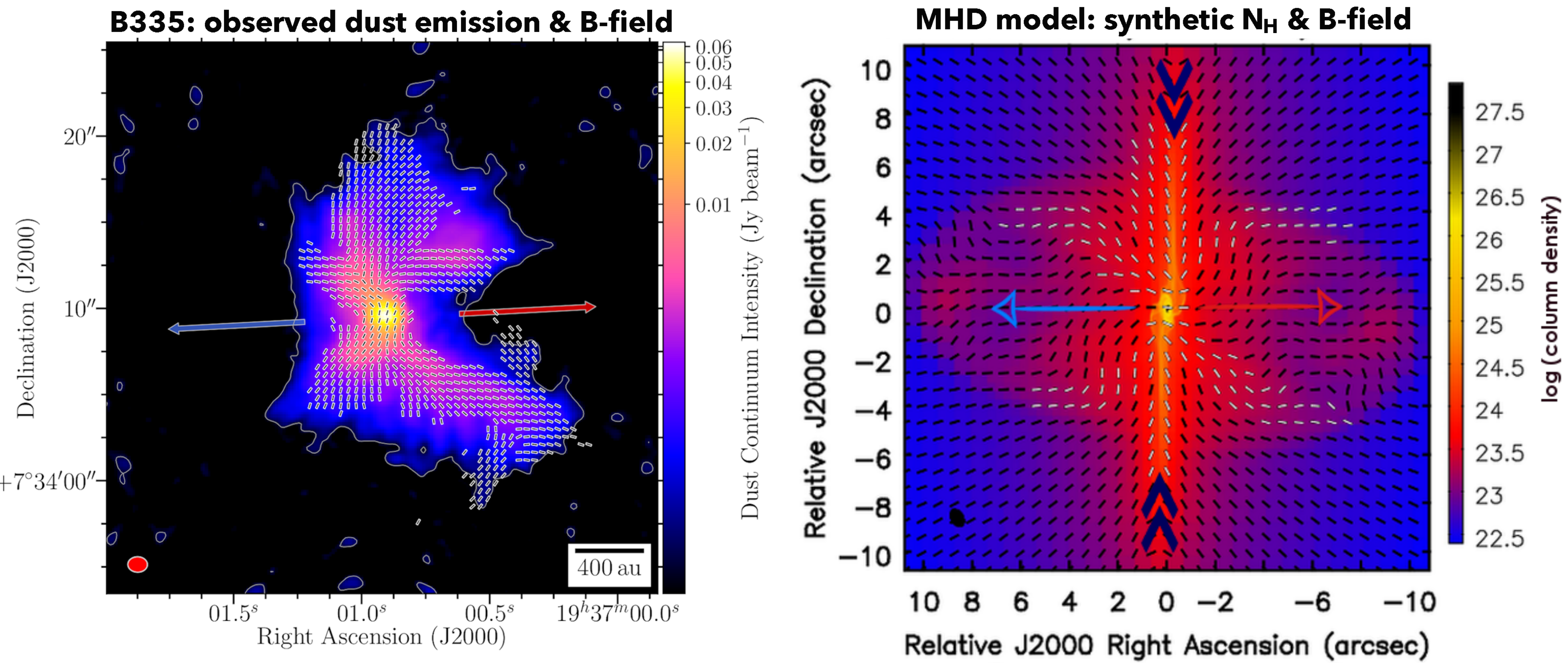}
\end{center}
\caption{Observation and best model to match the properties in the B335 protostar, from \citet{Maury18} and new ALMA observations (in prep.). The observed B field geometry shown in the left panel, the small disk size and the kinematics of the B335 inner envelope can only be explained by a family of MHD models where the initially poloidal field in B335 is being pulled in the dominant direction of the collapse, but yet is strong enough ($\mu \sim 6$) to counteract partially the transfer of angular momentum inwards and set the disk size in this Class 0 protostar: the best model is shown in the right panel.} 
\label{fig:B335}
\end{figure*}

The exact role of magnetic fields in setting the disks sizes and masses during the embedded phases can only be quantitatively addressed thanks to the comparison of observed properties to the outcomes of models. \citet{Maury18} have shown that the B field geometry, small disk size and the kinematics of the B335 inner envelope can only be explained by a family of MHD models where the initially poloidal field in B335 is being pulled in the dominant direction of the collapse, but yet is strong enough ($\mu \sim 6$) to counteract partially the transfer of angular momentum inwards. In B335, they show the magnetic field is very likely to regulate the formation processes of the protostellar disk, constraining the size of the protostellar disk to $< 20$ au (see Figure \ref{fig:B335}). The role of the magnetic field in setting the disk size  in B335 has also been discussed in subsequent studies \citep{Yen19,2019A&A...631A..64B,2019ApJ...873L..21I}. However, new observations by \citet{Cabedo22} suggest that the gas at small envelope radii is strongly ionized in B335, and points towards conditions typical of ideal MHD. It may very well be that the star-disk building phase consists in a succession of non-ideal and ideal MHD conditions in the ``life of the protostar". It would be possible to grow a quite large disk if it is formed before the onset of high ionization around the protostar (due to the protostar itself and the production of CR in the B-field lines around the protostar), while if the disks are still compact when the local ionization starts to increase at a fast pace, it may be more difficult to form large disks in the second half of the protostellar phase. 

\begin{figure*}[ht]
\begin{center}
\includegraphics[width=0.4\linewidth]{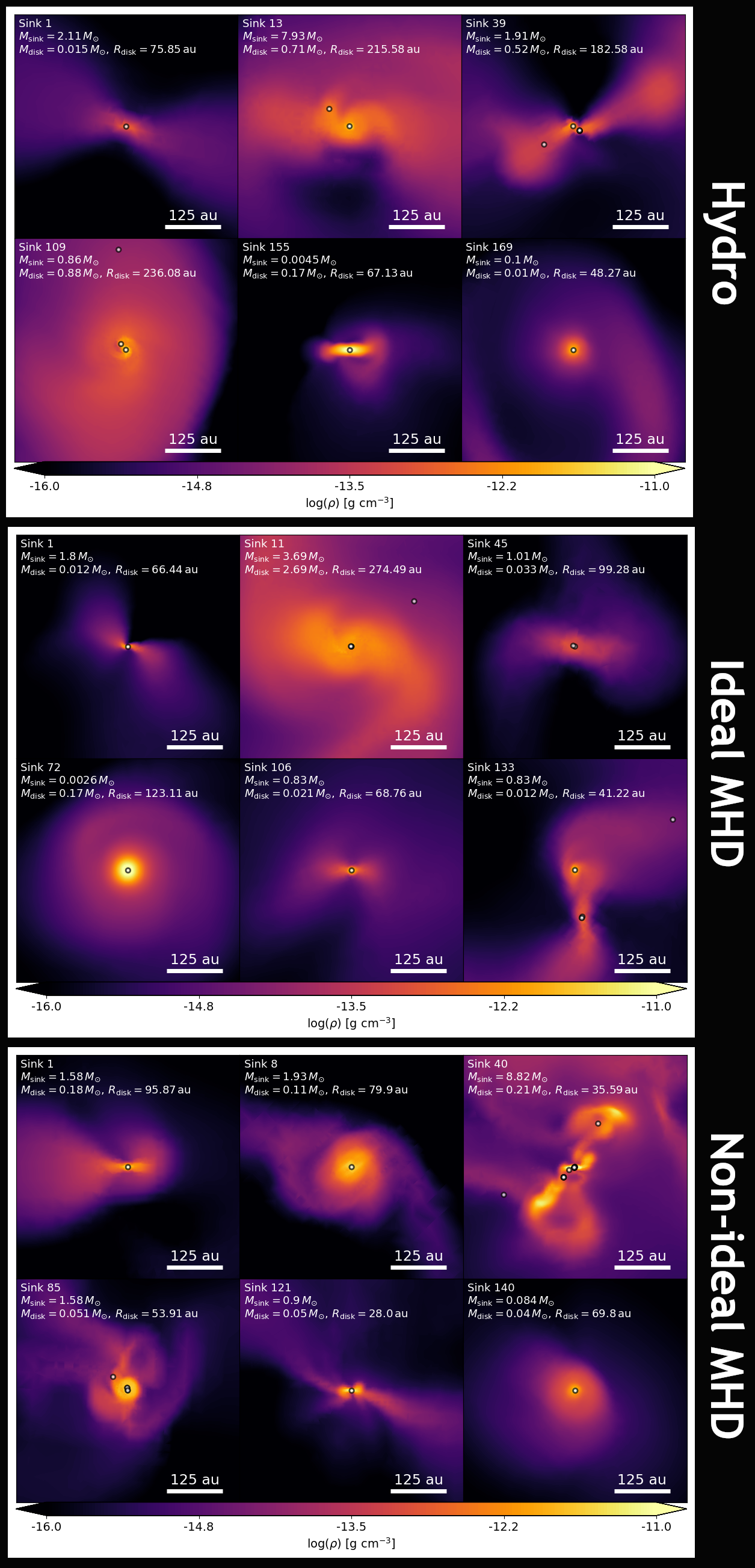}
\end{center}
\caption{\citet{lebreuilly2021}: A sample of disks formed in three massive clump simulations, hydrodynamics, ideal MHD 
and non-ideal MHD (ambipolar diffusion). Hydrodynamical disks tend to be bigger and ideal MHD ones smaller than when ambipolar 
diffusion is included. } 
\label{fig:lebreuilly-frag}
\end{figure*}

\subsection{Influence of magnetic field on the formation of multiple systems}

The fragmentation of low mass cores has been the subject of 
several studies 
\citep[see e.g.][]{matsumoto2003,goodwin04}.
Several modes of fragmentation have been identified.
Generally speaking, it is induced by the density 
fluctuations that are generated during the collapse
on one hand by gravo-turbulence processes within the envelope 
and on the other hand through the formation of massive centrifugally
supported unstable disks. 
 The outcome of
fragmentation therefore entirely depends on the initial conditions, particularly rotation and turbulence
initial values, but also on the amplitude of the initial
density perturbations. It has been generally found however
that under typical conditions and in the absence of 
magnetic field, 
a solar mass dense core tends to produce few fragments (say 2-10).

Several studies have been dedicated to the influence of magnetic field in this process
\citep[e.g.][]{machida2008,Hennebelle08b,commercon2010,wurster2017,wurster2019,hennebelle2020}, covering a broad 
range of parameters in terms of magnetic field intensity, 
rotation and initial turbulence. 
From these studies, clear trends can be inferred. 
First, the higher the rotational and/or turbulent energy, the more prone  to fragmentation is the core. Numerical models of turbulent dense cores suggest that the directions of the specific angular momentum axis tends to vary as a function of envelope scales \citep{joos2013,2017ApJ...839...69M}, and that misaligned systems can form in turbulent conditions. Some numerical models also suggest that turbulent velocity fluctuations around protostellar cores may be responsible for setting the rotational motions of these cores \citep{2019ApJ...881...11M}. However, magnetic field has a drastic influence and in the case 
of low mass cores, mass-to-flux ratios as high as 
$\mu > 5-10$ may entirely suppress the formation of fragments. 
This is because in low mass cores, turbulence is at best 
trans-sonic and the density fluctuations induced by 
turbulence remain limited. Therefore, the dominant 
fragmentation mode is through the formation of 
massive disks. However, in the presence of magnetic field 
the disk sizes and masses are much smaller, and therefore
they tend to be stable.
Another interesting trend is that the cores in which 
the magnetic field is initially perpendicular to the rotation 
axis are less prone to fragmentation than when it is aligned.  
One important fact to be taken into account however is that 
the amplitude of the density perturbations is also 
an important parameter, and the mentioned 
trends have been inferred for modest density fluctuations
(typically below 10$\%$). 
When the 
 density fluctuations have a  large amplitude,  say  of
about 50$\%$,  they are sufficiently unstable to collapse individually even in the absence of
rotation. In this case, magnetic field is unable to impede fragmentation \citep{Hennebelle08b,wurster2017}.
This would imply that the fragmentation is driven by 
the larger scales and the assembling process of the dense cores.

\begin{figure*}[ht]
\begin{center}
\includegraphics[width=0.7\linewidth]{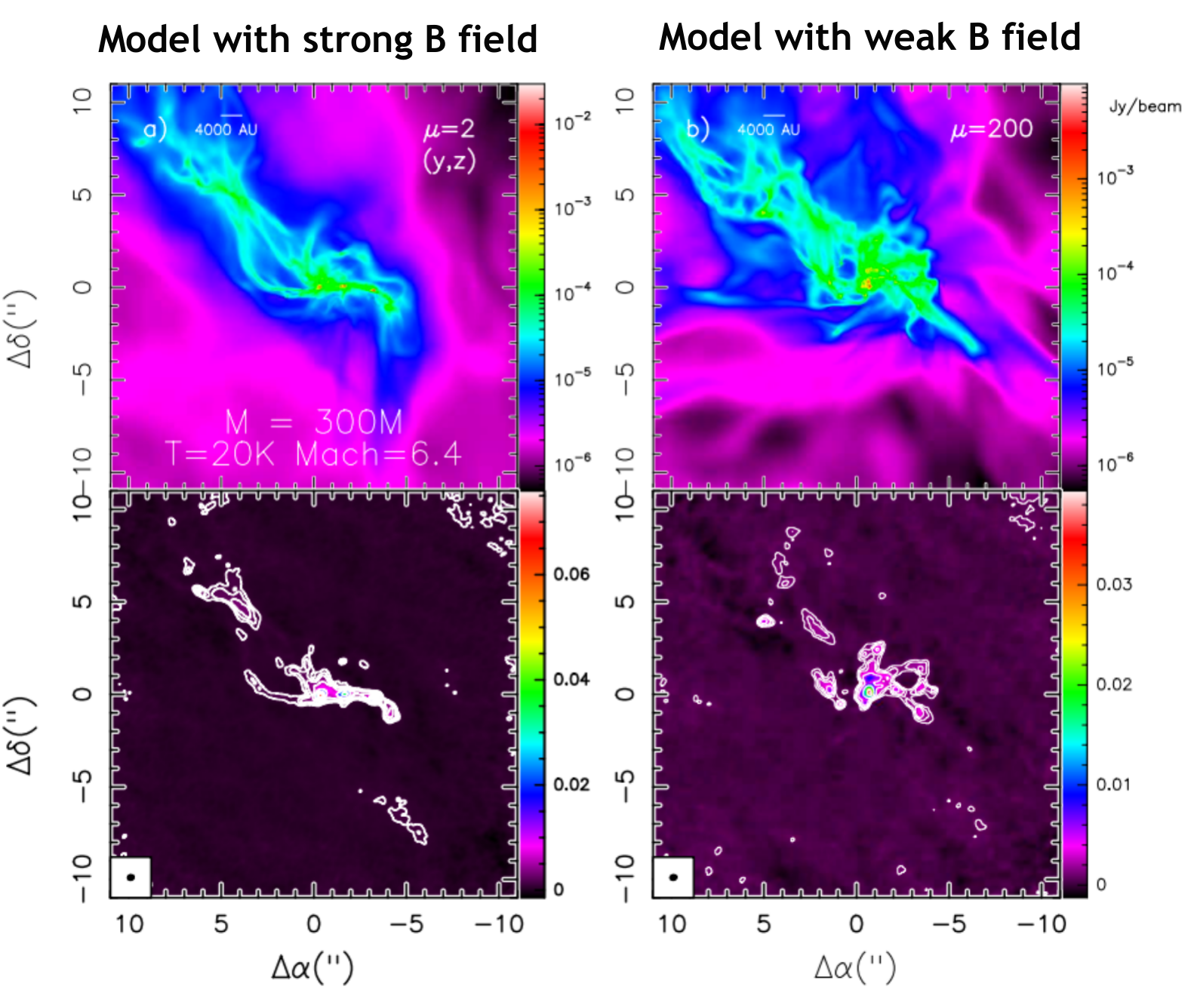}
\end{center}
\caption{Models of the gravitational collapse of a $300 \msun$ clump, with gas temperature T = 20 K, and Mach number 6.4, from \citet{fontani2018}. Column density maps of two models are shown in the upper row, in the presence of a strong (left) magnetic field and a weak one (right). The lower row shows ALMA synthetic observations of the models using the 280 GHz dust continuum emission. Fewer fragments are seen in the more magnetized case.}
\label{fig:fontani-frag}
\end{figure*}

The influence of the magnetic field on the fragmentation of massive clumps has also been investigated. Because of the 
large amplitude of the density fluctuations, the impact 
of magnetic field on the fragmentation is relatively 
less important than that found in the case of low 
mass cores. For instance, \citet{hennebelle2011}
and \citet{myers2013} found that in the presence of a significant magnetic field (e.g a few mG for mass-to-flux ratios around 2), the number of self-gravitating fragments is reduced by a factor of about 2. \citet{fontani2016} and \citet{fontani2018}
have carried out observations and simulations of massive clumps. 
They have studied and compared the  fragmentation properties of the clumps from observations and simulations using 
synthetic images that mimic the observations as 
shown in Fig.\ref{fig:fontani-frag}. They concluded 
that the fragmentation properties within the observations 
are better reproduced by the significantly magnetized models 
(typically $\mu=2$) than by the weakly magnetized ones.

\section{Summary}
\label{sec:conclusion}

Understanding how protostars and planet-forming disks form are cornerstone questions for us to understand both the stellar populations setting the evolution of our Universe and the conditions responsible for producing the planetary systems observed around most stars. The recent progresses made through observations and simulations suggest that magnetic field is modifying in depth the outcome of the collapse and the formation of stars, by adding an extra support to the gas, by reducing the angular momentum available to build the disks, by launching outflows 
and even possibly favoring the dust grain coagulation into pebbles. Observations suggest that the so-called magnetic braking catastrophe is an issue that is solved, alleviated by implementing realistic physical conditions (turbulence, anisotropies of the density field, non-homogeneous initial B field due to conditions of core assembly and local environment, gas ionization and dust properties) in MHD models. 

If the magnetized scenario we propose is common, recent work suggest that the angular momentum problem for star formation may be actually "solved" not by the formation of large disks but by the combination of (i) lack of organized rotation motions at large envelope radii, (ii) the inefficient angular momentum transport due to magnetic braking in the inner envelope (and angular momentum removed through rotating outflows generated by the presence of the magnetic field) and (iii) a local origin of the angular momentum incorporated in the star-disk system.

Major questions however remain to be solved for the next generations of astrophysicists. First, our understanding of the origin of the angular momentum that eventually build the disks is very incomplete. Whereas some of the gas momentum could obviously be inherited from larger scales, inertial processes naturally produce angular momentum in a non-axisymmetric system, opening the possibility to disks weakly related to large scale angular momentum. This possible scenario questions the effective role of magnetic braking at envelope scales to set the disk sizes, as local processes may be more important in shaping the outcome of star and disk formation. 
That means to better understand  the transport of the magnetic field by processes like reconnection diffusion and ambipolar diffusion and their importance on regulating angular momentum transport and disk formation in the different stages. Characterizing the properties of the compact protostellar disks, with very high angular resolution observations, will also be a key to set constraints and refine our models.
Second, this scenario deeply relies on the efficiency of magnetic fields to couple to the gas reservoirs. This efficiency is driven by a complex set of physical conditions, which remain up to this day largely unconstrained. Our current knowledge of magnetic resistivities remains thus hampered by very large uncertainties. It is therefore key, in the future, that observations characterize both the gas and dust properties in embedded protostars, and are compared to models so we know whether magnetic braking is much more efficient in locations, and epochs, that are critical for setting the disk and stellar properties.
Answering these questions will be a fascinating challenge that will require us to produce robust measurements of the magnetic intensity and topology in large samples of objects, but also find clever ways to access the gas ionisation fraction and dust grain properties in young embedded protostars.

\section*{Acknowledgments} AM acknowledges support from the European Research Council (ERC) under Starting grant number 679937 – `MagneticYSOs'. 
This work was also partially supported by the program Unidad de Excelencia María de Maeztu CEX2020-001058-M. JMG also acknowledges support by the grant PID2020-117710GB-I00 (MCI-AEI-FEDER,UE).
PH acknowledges support from the European Research Council
synergy grant ECOGAL (Grant : 855130)

\bibliographystyle{Frontiers-Harvard} 
\bibliography{bibreview.bib}

\end{document}